\pdfoutput=1 %
\documentclass[english
]{article} %

\usepackage{amsthm} %
\usepackage{array}
\usepackage{babel} %
\usepackage[ibidtracker=constrict,pagetracker=page,datamodel=software
,style=lncs
]{biblatex}
\usepackage{software-biblatex} %
\usepackage{fvextra} %
\usepackage{csquotes} %
\usepackage[inline]{enumitem}
\usepackage{fontawesome} %
\usepackage[T1]{fontenc} %
\usepackage{float} %
\usepackage{geometry}
\usepackage[colorlinks=true,allcolors=black]{hyperref} %
\usepackage{glossaries-extra} %
\usepackage[all]{hypcap} %
\usepackage[tt=false]{libertine} %
\usepackage{microtype}
\usepackage{myhyphenation} %
\usepackage[crefname,capitalize]{mymathpartir} %
\usepackage[bracketed=,macros]{mymathtools} %
\usepackage[outputdir=. %
,cachedir=minted                %
,finalizecache=false             %
,frozencache=true              %
]{minted}
\usepackage[
arrows
]{mystix2}
\usepackage[libertine]{newtxmath} %
\usepackage{tikz-cd}
\usepackage[disable]{todonotes}
\usepackage{zi4} %
\usepackage[capitalize,noabbrev,nameinlink]{cleveref} %
\usepackage{textgreek} %

\usepackage{xparse} %
\usepackage{xspace} %
\usepackage{zref}   %

\usepackage{myabbreviations} %
\usepackage{mycompounds}     %

\makeatletter
\NewDocumentCommand\dependson{mom}{%
  \@ifpackageloaded{#1}{%
    #3
  }{%
    \IfValueTF{#2}{
      \PackageWarningNoLine{macros}{Load package `#1' for #2}
    }{%
      \PackageWarningNoLine{macros}{Package `#1' needed for some macros but missing}
    }
  }
}
\NewDocumentCommand\dependsoncommand{mom}{%
  \ifcsname #1\endcsname
    #3
  \else
    \IfValueTF{#2}{
      \PackageWarningNoLine{macros}{Define command `#1' for #2}
    }{%
      \PackageWarningNoLine{macros}{Command `#1' needed for some macros but missing}
    }
  \fi
}
\makeatother

\ExplSyntaxOn
\NewDocumentCommand\MakeUppercaseFirst{m}{%
  \text_titlecase_first:n{#1}
}
\ExplSyntaxOff

\ProvideDocumentCommand{\grantsponsor}{mmm}{%
  #1%
}
\ProvideDocumentCommand{\grantnum}{omm}{%
  #3%
  \IfValueT{#1}{%
    \space(\url{#1})%
  }%
}
\def\acknowledgementsname{Acknowledgements}
\ifcsname acks\endcsname
\else
  \newenvironment{acks}{%
    \section*{\acknowledgementsname}
    \addcontentsline{toc}{section}{\acknowledgementsname}
  }{%
  }
\fi

\NewDocumentCommand{\thyeq}{}{\sim} %

\dependson{bbold}[\string\Bool,\string\Nat,\string\Three]{%
  \NewDocumentCommand{\Bool}{}{\mathbb{2}}
  \NewDocumentCommand{\Nat}{}{\mathbb{N}}
  \NewDocumentCommand{\Three}{}{\mathbb{3}}
}

\dependson{mathtools}[\string\Box]{%
  \dependson{amssymb}[\string\Box]{%
    \let\Box\relax\DeclareMathOperator{\Box}{\square}
  }
}

\dependson{mathtools}[\string\eval]{%
  \let\evallbracket\llbracket
  \let\evalrbracket\rrbracket
  \DeclarePairedDelimiter\eval\evallbracket\evalrbracket
}

\dependson{mathtools}[\string\pr, \string\copr]{%
  \DeclarePairedDelimiterX\pr[2]{\lparen}{\rparen}{#1,#2}
  \DeclarePairedDelimiterX\copr[2]{\lbrack}{\rbrack}{#1,#2}
}

\dependson{mathtools}[\string\Codiscr,\string\Discr,\string\Forget]{%
  \DeclareMathOperator{\Codiscr}{\nabla}
  \DeclareMathOperator{\Discr}{\Delta}
  \NewDocumentCommand{\Forget}{o}{%
    \IfNoValueTF{#1}{%
      \operatorname{U}
    }{%
      \operatorname{U}_{#1}
    }%
  }
}

\NewDocumentCommand{\newtheoremlist}{mo}{%
  \IfValueTF{#2}{%
    \newlist{#1list}{enumerate}[#2][#1]
  }{%
    \newlist{#1list}{enumerate}[#1][#1]
  }
}
\makeatletter
\let\orignewtheorem\newtheorem
\DeclareDocumentCommand{\newtheorem}{smomm}{%
  \IfBooleanTF{#1}{%
    \IfValueT{#3}{%
      \GenericWarning{%
        \string\newtheorem (macros)
      }{%
        Ignoring shared counter (#3) passed to starred version of
        \string\newtheorem
      }
    }
    \orignewtheorem*{#2}{#4}
    \newtheoremlist{#2}
  }{%
    \IfValueTF{#3}{%
      \orignewtheorem{#2}[#3]{#4}
      \newtheoremlist{#2}[#3]
    }{%
      \orignewtheorem{#2}{#4}
      \newtheoremlist{#2}
    }
  }
  \@ifpackageloaded{cleveref}{
    \crefname{#2}{#4}{#5}
    \Crefname{#2}{\expandafter\MakeUppercaseFirst{#4}}{\expandafter\MakeUppercaseFirst{#5}}
  }{}
}
\makeatother
\let\orignewacronym\newacronym
\DeclareDocumentCommand{\newacronym}{mmm}{%
  \dependson{glossaries-extra}{%
    \expandafter\NewDocumentCommand\expandafter{\csname#1\endcsname}{}{\glsfmtshort{#2}\xspace}
    \expandafter\NewDocumentCommand\expandafter{\csname#1s\endcsname}{}{\glsentryshortpl{#2}\xspace}
    \orignewacronym{#2}{#2}{#3}
  }
}
\makeatletter
\def\system@font{\rmfamily}
\def\systemfont{\system@font}
\NewDocumentCommand{\DeclareSystem}{mmm}{%
  \expandafter\DeclareRobustCommand\expandafter{\csname#1\endcsname}{{\system@font\glsfmttext{#1}}\xspace}
  \expandafter\DeclareRobustCommand\expandafter{\csname#1s\endcsname}{{\system@font\glsentryplural{#1}}\xspace}
  \newglossaryentry{#1}{%
    name={#2},
    description={#3}
  }
}
\makeatother

\NewDocumentCommand{\Agda}{}{\textsc{Agda}\xspace}

\NewDocumentCommand{\Haskell}{}{\textsc{Haskell}\xspace}
\NewDocumentCommand{\ML}{}{\textsc{ML}\xspace}
\NewDocumentCommand{\PureScript}{}{\textsc{PureScript}\xspace}

\newacronym{DOM}{DOM}{Document Object Model}
\newacronym{IFC}{IFC}{Information\hyp{}Flow Control}
\newacronym{NbE}{NbE}{Normalization by Evaluation}
\newacronym{OPE}{OPE}{Order\hyp{}preserving Embedding}

\makeatletter
\let\orignewlist\newlist
\def\label@enumerate{\arabic*.}
\def\ref@enumerate{\arabic*}
\DeclareRobustCommand{\perhaps}[1]{#1}
\NewDocumentCommand{\@itemref}{mm}{%
  \begingroup %
  \renewcommand\perhaps[1]{}%
  \csname #1ref\endcsname{#2}%
  \endgroup
}
\def\itemref{\@itemref{}}
\def\itemcref{\@itemref{c}}
\def\itemCref{\@itemref{C}}
\DeclareDocumentCommand\newlist{mmoo}{%
  \dependson{enumitem}{%
    \ifcsname label@#2\endcsname
      \ifcsname ref@#2\endcsname
        \orignewlist{#1}{#2}{1}
        \orignewlist{#1*}{#2*}{1}
        \IfValueTF{#3}{
          \setlist[#1,#1*]{label=\csname label@#2\endcsname,ref=\perhaps{\csname the#3\endcsname.}\csname ref@#2\endcsname}
        }{
          \setlist[#1,#1*]{label=\csname label@#2\endcsname.,ref=\csname ref@#2\endcsname}
        }
        \@ifpackageloaded{cleveref}{
          \IfValueTF{#4}{
            \crefalias{#1i}{#4}
            \crefalias{#1*i}{#4}
          }{
            \crefalias{#1i}{#1}
            \crefalias{#1*i}{#1}
          }
        }{}
      \else
        \GenericError{}{\string\newlist (macros)}{\string\ref@#2 undefined}{}
      \fi
    \else
      \GenericError{}{\string\newlist (macros)}{\string\label@#2 undefined}{}
    \fi
  }
}
\makeatother
\NewDocumentCommand{\NewTodo}{smm}{
  \expandafter\NewDocumentCommand\expandafter{\csname todo#2\endcsname}{om}{%
    \todo[\IfBooleanT{#1}{nolist},color=#3,\IfValueT{##1}{##1}]{}%
  }
}
\theoremstyle{plain}
\newtheorem{theorem}{Theorem}{Theorems}
\newtheorem{lemma}[theorem]{Lemma}{Lemmas}
\newtheorem{proposition}[theorem]{Proposition}{Propositions}
{Conjectures}
{Corollaries}
\theoremstyle{definition}
\newtheorem{definition}{Definition}{Definitions}
{Examples}
{Exercises}
{Questions}
\theoremstyle{remark}
{Answers}
{Remarks} 

\def\conjecturename{Conjecture}
\def\corollaryname{Corollary}
\def\definitionname{Definition}
\def\lemmaname{Lemma}
\def\propositionname{Proposition}
\def\theoremname{Theorem}
\makeatletter
\@ifpackageloaded{babel}{%
  \addto\extrasenglish{\renewcommand{\figurename}{Fig.}}
}{%
  \renewcommand{\figurename}{Fig.}
}
\makeatother

\NewTodo {error} {red!70}
\NewTodo*{warn}  {orange}
\NewTodo*{remark}{yellow!80!black}
\NewTodo*{hint}  {green!20}
\NewTodo*{done}  {blue!50}

\makeatletter
\newcounter{@inst}

\def\institutename{\par
 \begingroup
 \parskip=\z@
 \parindent=\z@
 \setcounter{@inst}{1}%
 \def\and{\par\stepcounter{@inst}%
 \noindent$^{\the@inst}$\enspace\ignorespaces}%
 \setbox0=\vbox{\def\thanks##1{}\@institute}%
 \ifnum\c@@inst=1\relax
   \gdef\fnnstart{0}%
 \else
   \xdef\fnnstart{\c@@inst}%
   \setcounter{@inst}{1}%
   \noindent$^{\the@inst}$\enspace
 \fi
 \ignorespaces
 \@institute\par
 \endgroup}

\def\inst#1{\unskip$^{#1}$}

\def\institute#1{%
  \gdef\@institute{#1}
  \date{\small\institutename}
}
\makeatother
\DeclareSymbolFont{sfoperators}{T1}{\sfdefault}{\mddefault}{\itdefault}

\makeatletter
\renewcommand{\operator@font}{\mathgroup\symsfoperators}
\makeatother

\renewcommand{\termfont}{\mathsf}

\crefname{subsection}{Subsection}{Subsections}
\Crefname{subsection}{Subsection}{Subsections}
\crefname{paragraph}{Paragraph}{Paragraphs}
\Crefname{paragraph}{Paragraph}{Paragraphs}

\crefname{corollary}{\corollaryname}{Corollaries}
\Crefname{corollary}{\corollaryname}{Corollaries}
\crefname{definition}{\definitionname}{\definitionname{}s}
\Crefname{definition}{\definitionname}{\definitionname{}s}
\crefname{example}{\examplename}{\examplename{}s}
\Crefname{example}{\examplename}{\examplename{}s}
\crefname{figure}{\figurename}{Figs.}
\Crefname{figure}{\figurename}{Figs.}
\crefname{lemma}{\lemmaname}{\lemmaname{}s}
\Crefname{lemma}{\lemmaname}{\lemmaname{}s}
\crefname{proof}{\proofname}{\proofname{}s}
\Crefname{proof}{\proofname}{\proofname{}s}
\crefname{proposition}{\propositionname}{\propositionname{}s}
\Crefname{proposition}{\propositionname}{\propositionname{}s}
\crefname{remark}{\remarkname}{\remarkname{}s}
\Crefname{remark}{\remarkname}{\remarkname{}s}
\crefname{theorem}{\theoremname}{\theoremname{}s}
\Crefname{theorem}{\theoremname}{\theoremname{}s}
\crefname{conjecture}{\conjecturename}{\conjecturename{}s}
\Crefname{conjecture}{\conjecturename}{\conjecturename{}s}

\DeclareMathOperator{\freshEnv}{freshEnv}
\DeclareMathOperator{\fresh}{fresh}
\DeclareMathOperator{\norm}{norm}

\DeclareMathOperator{\lookupTmS}{lookup}

\DeclareMathConstant{\Type}{Type}
\DeclareMathConstant{\StringSet}{String}

\let\quote=\relax
\DeclareMathOperator{\quote}{quote}
\DeclareMathOperator{\reify}{reify}
\DeclareMathOperator{\reflect}{reflect}

\DeclareMathConstant{\Ctx}{Ctx}
\DeclareMathConstant{\Ty}{Ty} %
\NewDocumentCommand{\Tm}{mm}{#1 \vdash #2} %
\NewDocumentCommand{\Ne}{mm}{#1 \vdashNe #2} %
\NewDocumentCommand{\Nf}{mm}{#1 \vdashNf #2} %
\NewDocumentCommand{\Ri}{}{\mathrel{R_i}}
\DeclareMathOperator{\reflRi}{refl_{i}}
\DeclareMathOperator{\transRi}{trans_{i}}
\NewDocumentCommand{\Rm}{}{\mathrel{R_m}}
\DeclareMathOperator{\reflRm}{refl_{m}}
\DeclareMathOperator{\transRm}{trans_{m}}
\DeclareMathOperator{\factorRi}{factor_{i}}
\DeclareMathOperator{\factorRm}{factor_{m}}

\DeclareMathOperator{\LR}{L}

\DeclareTypeInfix{\FunTy}{\Rightarrow}
\DeclareTypeInfix{\ProdTy}{\times}
\NewDocumentCommand{\BoxTy}{g}{\square\IfValueT{#1}{#1}} %
\DeclareTypeConstant{\BaseTy}{\iota}
\DeclareTypeConstant{\UnitTy}{Unit}
\DeclareTypeConstant{\BoolTy}{Bool}
\DeclareTypeConstant{\NatTy}{N}
\NewDocumentCommand{\lockTy}{g}{\blacklozenge\IfValueT{#1}{#1}} %
\NewDocumentCommand{\NegTy}{g}{\IfValueTF{#1}{\neg#1}{\neg}} %

\NewDocumentCommand{\Wk}{mm}{#1 \leq #2}
\def\cesymb{\vartriangleleft}
\NewDocumentCommand{\R}{mm}{#1 \cesymb #2}
\NewDocumentCommand{\RIKC}{mm}{#1 \cesymb_{\text{\IKC}} #2}
\NewDocumentCommand{\RISFourC}{mm}{#1 \cesymb_{\text{\ISFourC}} #2}
\NewDocumentCommand{\RIKFourC}{mm}{#1 \cesymb_{\text{\IKFourC}} #2}
\NewDocumentCommand{\RITC}{mm}{#1 \cesymb_{\text{\ITC}} #2}

\DeclareTermConstant{\EmptyCtx}{{\cdot}}
\NewDocumentCommand{\ExtCtx}{mm}{#1, #2}
\DeclareTypeConstant{\lockCtx}{\textnormal{\faLock}}

\NewDocumentCommand{\ProdCtx}{mm}{#1, #2}

\DeclareTermConstant{\baseWk}{base}
\DeclareTermOperator{\dropWk}{drop}
\DeclareTermOperator{\keepVarWk}{keep}
\DeclareTermOperator{\keepLockWk}{keep_{\lockCtx}}
\DeclareMathOperator{\wk}{wk}
\DeclareMathOperator{\leftConcat}{leftConcat}
\DeclareMathOperator{\factor}{factor}
\DeclareTermConstant{\idWk}{id_{\leq}}
\DeclareTermConstant{\dropIdWk}{\dropWk{\idWk}}

\DeclareTermOperator{\varExt}{var}
\DeclareTermOperator{\nilExt}{nil}
\DeclareTermOperator{\lockExt}{lock}

\DeclareMathOperator{\toOPE}{toOPE}

\DeclareTermOperator{\emptySub}{empty}
\DeclareTermOperator{\extTmSub}{ext}
\DeclareTermOperator{\extLockSub}{ext_{\lockCtx}}
\DeclareTermOperator{\lockSub}{lock}
\DeclareTermConstant{\idSub}{id_{\textsc{s}}}
\DeclareMathOperator{\trimIKSubS}    {trim_{\text{\IKC}}}     %
\DeclareMathOperator{\trimISFourSubS}{trim_{\text{\ISFourC}}} %
\DeclareMathOperator{\trimITSubS}    {trim_{\text{\ITC}}}     %
\DeclareMathOperator{\trimIKFourSubS}{trim_{\text{\IKFourC}}} %
\DeclareMathOperator{\subst}{subst}

\NewDocumentCommand{\vdashVar}{}{\vdash_{\textsc{var}}}
\NewDocumentCommand{\vdashNf}{}{\vdash_{\textsc{nf}}}
\NewDocumentCommand{\vdashNe}{}{\vdash_{\textsc{ne}}}
\NewDocumentCommand{\vdashSub}{}{\vdash_{\textsc{s}}}

\NewDocumentCommand{\stepsto}{}{\mapsto}

\DeclareTermConstant{\zeroVar}{zero}
\DeclareTermConstant{\unitTm}{unit}
\DeclareTermConstant{\trueTm}{true}
\DeclareTermConstant{\falseTm}{false}

\DeclareTermOperator{\succVar}{succ}
\DeclareTermOperator{\varTm}{var}
\DeclareTermOperator{\diaTm}{dia}
\DeclareTermOperator{\boxTm}{box}
\DeclareTermOperator{\unboxTm}      {unbox}
\DeclareTermOperator{\unboxIKTm}    {unbox_{\text{\IKC}}}
\DeclareTermOperator{\unboxISFourTm}{unbox_{\text{\ISFourC}}}
\DeclareTermOperator{\unboxITTm}    {unbox_{\text{\ITC}}}
\DeclareTermOperator{\unboxIKFourTm}{unbox_{\text{\IKFourC}}}
\DeclareTermOperator{\lamTm}{\lambda}
\DeclareTermOperator{\appTm}{app}
\DeclareTermOperator{\upTm}{up}
\DeclareTermOperator{\ifteTm}{ifte}
\DeclareTermInfix{\andTm}{\wedge}
\NewDocumentCommand{\negTm}{g}{\neg\IfValueT{#1}{#1}} %
\DeclareTermOperator{\bindTm}{bind}
\DeclareTermOperator{\returnTm}{return}
\DeclareTermOperator{\liftTm}{lift}
\DeclareTermInfix{\mulTm}{\ast}

\DeclareSystem{IK}    {IK}  {Intuitionistic \KA}
\DeclareSystem{IKFour}{IK4} {\IK with \FourA}
\DeclareSystem{ISFour}{IS4} {\IK with \SA and \FourA}
\DeclareSystem{IT}    {IT}  {\IK with \TA}
\DeclareSystem{IR}    {IR}  {\IK with \RA}
\DeclareSystem{STLC}  {STLC}{Simply\hyp{}typed lambda calculus}
\DeclareSystem{MTT}   {MTT} {Multimodal Type Theory}
\DeclareSystem{CCC}   {CCC} {Cartesian\hyp{}closed category}

\DeclareSystem{IKC}    {\protect\textnormal{\textlambda}\textsubscript{\IK}}{}
\DeclareSystem{IKFourC}{\protect\textnormal{\textlambda}\textsubscript{\IKFour}}{}
\DeclareSystem{ISFourC}{\protect\textnormal{\textlambda}\textsubscript{\ISFour}}{}
\DeclareSystem{ITC}    {\protect\textnormal{\textlambda}\textsubscript{\IT}}{}
\DeclareSystem{IRC}    {\protect\textnormal{\textlambda}\textsubscript{\IR}}{}

\DeclareSystem{MoggiCap}{Moggi\textsuperscript{Cap}}{}

\DeclareSystem{DCC}{DCC}{Dependency Core Calculus}
\DeclareSystem{MAC}{MAC}{Static Mandatory Access Control in \Haskell}

\def\KA   {\TirName{K}\xspace}
\def\SA   {\TirName{S}\xspace}
\def\TA   {\TirName{T}\xspace}
\def\RA   {\TirName{R}\xspace}
\def\FourA{\TirName{4}\xspace}
\def\GLA  {\TirName{GL}\xspace}

\DeclareSystem{IKCBool}{\protect\textnormal{\textlambda}\textsubscript{\IK}+{$\BoolTy$}}{}
\DeclareSystem{IKCNat}{\protect\textnormal{\textlambda}\textsubscript{\IK}+{$\NatTy$}}{}
\DeclareSystem{ISFourCBool}{\protect\textnormal{\textlambda}\textsubscript{\ISFour}+{$\BoolTy$}}{}
\DeclareSystem{ISFourCMoggiCap}{\protect\textnormal{\textlambda}\textsubscript{\ISFour}+\MoggiCap}{}

\DeclareTermOperator{\printTm}{print}
\DeclareTermOperator{\letTm}{let}
\NewDocumentCommand{\strTm}{m}{\termfont{str}_{#1}}

\DeclareTermOperator{\unsafePerformIOTm}{unsafePerformIO}

\DeclareTypeConstant{\CapTy}{Cap}
\DeclareTypeConstant{\StringTy}{String}

\DeclareTypeOperator{\IOTy}{IO}
\DeclareTypeOperator{\TTy}{T}

\begin{document}
\title{Normalization for Fitch\texorpdfstring{\hyp{}}{-}Style Modal Calculi}

\author{Nachiappan Valliappan\inst{1} \and
Fabian Ruch \and
Carlos Tomé Cortiñas\inst{1}}
\makeatletter
\institute{Chalmers University of Technology\stepcounter{@inst}} %
\makeatother
\maketitle %

\begin{abstract}
  Fitch\hyp{}style modal lambda calculi enable programming with
  necessity modalities in a typed lambda calculus by extending the typing
  context with a delimiting operator that is denoted by a lock.
  The addition of locks simplifies the formulation of typing rules
  for calculi that incorporate different
  modal axioms, but each variant demands different,
  tedious and seemingly ad~hoc syntactic lemmas
  to prove normalization.
  In this work, we take a semantic approach to normalization, called normalization
  by evaluation (\NbE), by leveraging the possible\hyp{}world semantics of Fitch\hyp{}style calculi to yield a more modular approach to normalization.
  We show that \NbE models can be constructed for calculi that incorporate the \KA, \TA
  and \FourA axioms of modal logic, as suitable instantiations of the possible\hyp{}world
  semantics.
  In addition to existing results that handle $\beta$\nbhyp{}equivalence,
  our normalization result also considers $\eta$\nbhyp{}equivalence for these calculi.
  Our key results have been mechanized in the proof assistant \Agda. Finally,
  we showcase several
  consequences of normalization
  for proving meta\hyp{}theoretic properties of Fitch\hyp{}style calculi
  as well as programming\hyp{}language applications based on different interpretations of the
  necessity modality.
\end{abstract}

\section{Introduction}\label{sec:introduction}

In type systems, a \emph{modality} can be broadly construed as a
unary type constructor with certain properties.
Type systems with modalities have found a wide range of applications
in programming languages to specify properties of a program in its
type.
In this work, we study typed lambda calculi equipped
with a \emph{necessity} modality (denoted by $\BoxTy$) formulated
in the so\hyp{}called Fitch style.

The necessity modality originates from modal logic, where the most basic
intuitionistic modal logic~\IK (for ``intuitionistic'' and ``Kripke'')
extends intuitionistic propositional logic with a unary connective $\BoxTy$,
the \emph{necessitation
  rule} (if $\EmptyCtx \vdash A$ then $\Gamma \vdash \BoxTy{A}$) and the
\emph{\KA axiom}
($\BoxTy{(A \FunTy B)} \FunTy \BoxTy{A} \FunTy \BoxTy{B}$).
With the addition of further modal axioms~\TA
($\BoxTy{A} \FunTy A$) and~\FourA ($\BoxTy{A} \FunTy \BoxTy{\BoxTy{A}}$)
to \IK, we obtain richer logics~\IT (adding axiom~\TA),~\IKFour
(adding axiom~\FourA),
and~\ISFour (adding both \TA and \FourA).
Type systems with necessity modalities based on \IK and \ISFour
have found applications in partial evaluation and staged computation~\parencite{DaviesP96,DaviesP01},
information\hyp{}flow control~\parencite{MiyamotoI04}, and recovering purity in
an effectful language~\parencite{ChoudhuryK20}.
While type systems based on \IT and \IKFour do not seem to have
any prior known programming applications, they are nevertheless interesting
as objects of study that extend \IK towards \ISFour.

Fitch\hyp{}style modal lambda calculi~\parencite{Borghuis94,Clouston18,MartiniM96}
feature necessity modalities in a typed lambda
calculus by extending the typing context
with a delimiting ``lock'' operator (denoted by $\lockCtx$).
In this paper, we consider the family of Fitch\hyp{}style modal lambda calculi
that correspond to the logics~\IK,~\IT,~\IKFour, and~\ISFour.
These calculi extend the simply\hyp{}typed lambda
calculus~(\STLC) with a type constructor~$\BoxTy$, along with introduction
and elimination rules for $\BoxTy$ types formulated
using the $\lockCtx$ operator.
For instance, the calculus~\IKC, which corresponds to the logic~\IK,
extends \STLC with~\cref{rule:boxTm-0,rule:unboxTm/IKC-0},
as summarized in \cref{fig:ikc-syn}.
The rules for $\lambda$\nbhyp{}abstraction and
function application are formulated in the usual way\mdash{}but note
the modified variable rule~\labelcref{rule:varTm-0}!

\begin{figure}[ht]
  \begin{align*}
    \Ty\qquad A \Coloneqq \ldots\ |\ \BoxTy{A} & &
    \Ctx\qquad \Gamma \Coloneqq \EmptyCtx\ |\ \ExtCtx{\Gamma}{x : A}\ |\ \ExtCtx{\Gamma}{\lockCtx}
  \end{align*}
  \begin{mathpar}
    \inferrule[Var]{ }
    [\lockCtx \notin \Gamma']{%
      \ProdCtx{\ExtCtx{\Gamma}{x : A}}{\Gamma'} \vdash x : A
    }\label{rule:varTm-0}

    \inferrule[$\BoxTy$\nbhyp{}Intro]{%
      \ExtCtx{\Gamma}{\lockCtx} \vdash t : A
    }{%
      \Gamma \vdash \boxTm{t} : \BoxTy{A}
    }\label{rule:boxTm-0}

    \inferrule[\IKC/$\BoxTy$\nbhyp{}Elim]{%
      \Gamma  \vdash t : \BoxTy{A}
    }[\lockCtx \notin \Gamma']{%
      \ProdCtx{\ExtCtx{\Gamma}{\lockCtx}}{\Gamma'} \vdash \unboxIKTm{t} : A
    }\label{rule:unboxTm/IKC-0}
  \end{mathpar}
  \caption{Typing rules for \IKC (omitting $\lambda$\nbhyp{}abstraction and application)}
  \label{fig:ikc-syn}
\end{figure}

The equivalence of terms in \STLC is extended by
Fitch\hyp{}style calculi with the following rules for $\BoxTy$ types,
where the former states the $\beta$\nbhyp{} (or computational) equivalence,
and the latter states a type\hyp{}directed $\eta$\nbhyp{}
(or extensional) equivalence.
\begin{mathpar}
\inferrule[$\BoxTy$\nbhyp{}$\beta$]
  {}
  {\unboxTm{(\boxTm{t})} \thyeq t}\label{rule:box-beta}

\inferrule[$\BoxTy$\nbhyp{}$\eta$]
  {\Gamma \vdash t : \BoxTy{A}}
  {t \thyeq \boxTm{(\unboxTm{t})}}\label{rule:box-eta}
\end{mathpar}
We are interested in the problem of normalizing terms
with respect to these equivalences.
Traditionally, terms in a calculus are normalized by rewriting
them using rewrite rules formulated from these equivalences, and a term
is said to be in \emph{normal form} when it cannot be rewritten further.
For example, we may formulate a rewrite rule
$\unboxTm{(\boxTm{t})} \stepsto t$ by orienting the \labelcref{rule:box-beta} equivalence
from left to right.
This naive approach to formulating a rewrite rule,
however, is insufficient for the \labelcref{rule:box-eta} rule since
normalizing with a rewrite rule
$t \stepsto \boxTm{(\unboxTm{t})}$ (for $\Gamma \vdash t : \BoxTy{A}$) does not terminate
as it can be applied infinitely many times.
It is presumably for this reason that existing normalization results~\parencite{Clouston18}
for some of these calculi only consider $\beta$\nbhyp{}equivalence.

While it may be possible to carefully formulate a more complex set of
rewrite rules that take the context of application into consideration
to guarantee termination (as done, for example, by \textcite{JayG95} for
function and product types), the situation is further complicated for
Fitch\hyp{}style calculi by the fact that we must repeat such syntactic
rewriting arguments separately for each calculus under consideration.
The calculi~\ITC,~\IKFourC, and~\ISFourC differ from \IKC
only in the $\BoxTy$\nbhyp{}elimination rule, as
summarized in \cref{fig:it-ik4-is4-syn}.
\begin{figure}[ht]
  \begin{mathpar}
    \inferrule[\ITC/$\BoxTy$\nbhyp{}Elim]{%
      \Gamma  \vdash t : \BoxTy{A}
    }[\#_\lockCtx(\Gamma') \leq 1]{%
      \ProdCtx{\Gamma}{\Gamma'} \vdash \unboxITTm{t} : A
    }\label{rule:unboxTm/ITC-0}

    \inferrule[\IKFourC/$\BoxTy$\nbhyp{}Elim]{%
      \Gamma  \vdash t : \BoxTy{A}
    }{%
      \ProdCtx{\ExtCtx{\Gamma}{\lockCtx}}{\Gamma'} \vdash \unboxIKFourTm{t} : A
    }\label{rule:unboxTm/IKFourC-0}

    \inferrule[\ISFourC/$\BoxTy$\nbhyp{}Elim]{%
      \Gamma  \vdash t : \BoxTy{A}
    }{%
      \ProdCtx{\Gamma}{\Gamma'} \vdash \unboxISFourTm{t} : A
    }\label{rule:unboxTm/ISFourC-0}
  \end{mathpar}
  \caption{$\BoxTy$\nbhyp{}elimination rules for \ITC, \IKFourC, and \ISFourC}
  \label{fig:it-ik4-is4-syn}
\end{figure}
In spite of having identical syntax and term equivalences,
each calculus demands different, tedious and seemingly ad~hoc
syntactic renaming lemmas~\parencite[Lemmas 4.1 and 5.1]{Clouston18}
to prove normalization.

In this paper, we take a semantic approach to normalization, called
normalization by evaluation~(\NbE)~\parencite{BergerS91}.
\NbE bypasses rewriting entirely, and instead normalizes terms by
evaluating them in a suitable semantic model and then reifying
values in the model as normal forms.
For Fitch\hyp{}style calculi, \NbE can be developed by leveraging their
possible\hyp{}world semantics.
To this end, we identify the parameters of the possible\hyp{}world semantics
for the calculi under consideration, and show that \NbE models
can be constructed by instantiating those parameters.
The \NbE approach exploits the semantic overlap of the Fitch\hyp{}style
calculi in the possible\hyp{}world semantics and isolates their differences
to a specific parameter that determines the modal fragment, thus
enabling the reuse of the evaluation machinery and many lemmas proved in the process.

In \cref{sec:main-idea}, we begin by providing a brief overview of the
main idea underlying this paper.
We discuss the uniform interpretation of types for four Fitch\hyp{}style calculi
(\IKC, \ITC, \IKFourC and \ISFourC) in possible\hyp{}world models
and outline how \NbE models can be constructed as instances.
The \emph{reification} mechanism that enables \NbE is performed alike
for all four calculi.
In \cref{sec:nbe-models}, we construct an \NbE model
for \IKC that yields a correct
normalization algorithm, and then show how \NbE models can
also be constructed for \ISFourC, and for \ITC and \IKFourC
by slightly varying the instantiation%
.
The calculi~\IKC and~\ISFourC and their normalization algorithms
have been implemented and verified correct~\todoerror{}{\parencite{ValliappanRT22v1.0.2}}
in the proof assistant~\Agda~\parencite{Agda2v2.6.2.1}.

\NbE models and proofs of normalization in
general have several useful consequences for term calculi.
In \cref{sec:corollaries}, we show how \NbE models
and the accompanying normalization algorithm
can be used to prove meta\hyp{}theoretic properties
of Fitch\hyp{}style calculi including completeness,
decidability, and some standard results in modal logic
in a \emph{constructive} manner.
In \cref{sec:applications}, we discuss applications of our
development to specific interpretations of the necessity modality in
programming languages, and show (but do not mechanize) how
application\hyp{}specific properties that typically require semantic
intervention can be proved syntactically.
We show that properties similar to capability safety, noninterference,
and binding\hyp{}time correctness can be proved syntactically using
normal forms of terms.

\section{Main Idea}\label{sec:main-idea}

The main idea underlying this paper is that normalization can be achieved
in a modular fashion for Fitch\hyp{}style calculi by constructing
\NbE models as instances of their possible\hyp{}world semantics.
In this \lcnamecref{sec:main-idea}, we observe that Fitch\hyp{}style
calculi can be interpreted in the possible\hyp{}world semantics for
intuitionistic modal logic with a minor refinement that accommodates
the $\lockCtx$ operator, and give a brief overview of how we
construct \NbE models as instances.

\paragraph{Possible\texorpdfstring{\hyp{}}{-}World Semantics}\label{para:possible-world}
The possible\hyp{}world semantics for intuitionistic modal
logic~\parencite{BozicD84} is parameterized by a \emph{frame}~$F$ and
a \emph{valuation}~$V_\BaseTy$.
A frame~$F$ is a triple~$\triple{W}{\Ri}{\Rm}$ that consists
of a type~$W$ of \emph{worlds} along with two binary
\emph{accessibility} relations~$\Ri$ (for ``intuitionistic'')
and~$\Rm$ (for ``modal'') on worlds that are required to satisfy
certain conditions.
An element~$w : W$ can be thought of as a representation of the
``knowledge state'' about some ``possible world'' at a certain point
in time.
Then, $w \Ri w'$ represents an increase in knowledge from $w$ to $w'$,
and $w \Rm v$ represents a possible passage from $w$ to $v$.
A valuation~$V_{\BaseTy}$, on the other hand, is a family of
types~$V_{\BaseTy,w}$ indexed by $w : W$ along with
functions~$\wk_{\BaseTy,w,w'} : V_{\BaseTy,w} \to V_{\BaseTy,w'}$
whenever $w \Ri w'$.
An element~$p : V_{\BaseTy,w}$ can be thought of as ``evidence'' for
(the knowledge of) the truth of the \emph{atomic}
proposition~$\BaseTy$ at the world~$w$.
The requirement for functions~$\wk_{\BaseTy,w,w'}$
enforces that the knowledge of the truth of $\BaseTy$ at $w$
is preserved as time moves on to $w'$, and is neither
forgotten nor contradicted by any new evidence learned at $w'$.
There are no such requirements on a valuation~$V_{\BaseTy}$ with
respect to the modal accessibility relation~$\Rm$.

Given a frame~$\triple{W}{\Ri}{\Rm}$ and a valuation~$V_{\BaseTy}$,
we interpret (object) types~$A$ in \emph{any} Fitch\hyp{}style
calculus as families of (meta) types~$\eval{A}_{w}$ indexed by
worlds~$w : W$,
following the work
by \textcite{Fischer-Servi81,Ewald86,PlotkinS86,Simpson94a} as below:
\begin{equation*}
\begin{array}{>{\evallbracket}l@{\evalrbracket}l @{\;}c@{\;} l}
  \BaseTy    & _w & = & V_{\BaseTy,w} \\
  A \FunTy B & _w & = & \forall w'.\, w \Ri w' \to \eval{A}_{w'} \to \eval{B}_{w'} \\
  \BoxTy{A}  & _w & = & \forall w'.\, w \Ri w' \to \forall v.\, w' \Rm v  \to \eval{A}_{v}
\end{array}
\end{equation*}

The nonmodal type formers are interpreted as in the Kripke semantics
for intuitionistic propositional logic: the base type~$\BaseTy$ is interpreted using
the valuation~$V_\BaseTy$, and function types~$A \FunTy B$ at $w : W$
are interpreted as \emph{families} of
functions~$\eval{A}_{w'} \to \eval{B}_{w'}$ indexed by $w' : W$ such
that $w \Ri w'$.
Recall that the generalization to families is necessary for the
interpretation of function types to be sound%
.

As for the interpretation of modal types, at $w : W$ the
types~$\BoxTy{A}$ are interpreted by families of
elements~$\eval{A}_{v}$ indexed by those $v : W$ that are accessible
from $w$ via some $w' : W$ such that $w \Ri w'$ and $w' \Rm v$.
In other words, $\BoxTy{A}$ is true at a world~$w$ if $A$ is
necessarily true in ``the future'', whichever concrete possibility
this may turn out to be.
We remark that the interpretation of $\BoxTy{A}$ as $\forall
v.\, w \Rm v \to \eval{A}_{v}$, as in classical modal logic without the first
quantifier~$\forall w'.\, w \Ri w'$, requires additional conditions~\parencite{BozicD84,Simpson94a}
on frames that (some of) the
\NbE models we construct do not satisfy.

In order to extend the possible\hyp{}world semantics of intuitionistic
modal logic to Fitch\hyp{}style calculi, we must also provide an
interpretation of contexts and the $\lockCtx$ operator, which is
unique to the Fitch style, in particular:
\begin{equation*}
\begin{array}{>{\evallbracket}l@{\evalrbracket}l @{\;}c@{\;} l}
  \EmptyCtx                 & _w & = & \top \\
  \ExtCtx{\Gamma}{A}        & _w & = & \eval{\Gamma}_w \times \eval{A}_w \\
  \ExtCtx{\Gamma}{\lockCtx} & _w & = & \textstyle\sum\nolimits_{u} {\eval{\Gamma}_u \times u \Rm w}
\end{array}
\end{equation*}

The empty context~$\EmptyCtx$ and the context
extension~$\ExtCtx{\Gamma}{A}$ of a context~$\Gamma$ with a type~$A$
are interpreted as in the Kripke semantics for \STLC by the terminal
family and the Cartesian product of the families~$\eval{\Gamma}$
and~$\eval{A}$, respectively.
While the interpretation of types~$\BoxTy{A}$ can be understood as a
statement about the future, the interpretation of
contexts~$\ExtCtx{\Gamma}{\lockCtx}$ can be understood as a dual
statement about the past: $\ExtCtx{\Gamma}{\lockCtx}$ is true at a
world~$w$ if $\Gamma$ is true at \emph{some} world~$u$ for which $w$
is a possibility, \ie $u \Rm w$.

With the interpretation of contexts~$\Gamma$ and types~$A$ as
$\pair{W}{\Ri}$\nbhyp{}indexed families~$\eval{\Gamma}$ and~$\eval{A}$
at hand, the interpretation of terms~$t : \Tm{\Gamma}{A}$, also known
as \emph{evaluation}, in a possible\hyp{}world model is given by a
function~\todowarn{}{$\eval : \Tm{\Gamma}{A} \to (\forall w .\, \eval{\Gamma}_w
  \to \eval{A}_w)$} as follows.
\Textcite{Clouston18} shows that the interpretation of \STLC in
Cartesian closed categories~(\CCCs) extends to an interpretation of
Fitch\hyp{}style calculi in any \CCC equipped with an adjunction
by interpreting $\BoxTy$ and $\lockCtx$ by the right and left adjoint
as well as $\boxTm$ and $\unboxTm$ using the right and left adjuncts,
respectively.
The key idea here is that, correspondingly, the interpretation of
terms in the nonmodal fragment of Fitch\hyp{}style calculi using the
familiar \CCC structure on $\pair{W}{\Ri}$\nbhyp{}indexed families
extends to the modal fragment: the interpretation of $\BoxTy$ in a
possible\hyp{}world model has a left adjoint that is denoted by our
interpretation of $\lockCtx$.
In summary, the possible\hyp{}world interpretation of Fitch\hyp{}style
calculi can be given by instantiation of \citeauthor{Clouston18}'s
\emph{generic} interpretation in \CCCs equipped with an adjunction.

\paragraph{Constructing \texorpdfstring{\NbE}{NbE} Models as Instances}
To construct an \NbE model for Fitch\hyp{}style calculi, we must construct a
possible\hyp{}world model with a
function~$\quote : (\forall w .\, \eval{\Gamma}_w \to \eval{A}_w) \to \Gamma \vdashNf A$
that inverts the denotation~$(\forall w .\, \eval{\Gamma}_w \to \eval{A}_w)$ of a term
to a derivation~$\Gamma \vdashNf A$ in normal form.
The normal forms for the modal fragment of \IKC are defined below,
where $\Gamma \vdashNe A$ denotes a special case of normal forms
known as \emph{neutral elements}.
\begin{mathpar}
  \inferrule[Nf/$\BoxTy$\nbhyp{}Intro]{%
    \ExtCtx{\Gamma}{\lockCtx} \vdashNf t : A
  }{%
    \Gamma \vdashNf \boxTm{t} : \BoxTy{A}
  }\label{rule:boxNf-0}%

  \inferrule[\IKC/Ne/$\BoxTy$\nbhyp{}Elim]{%
    \Gamma  \vdashNe t : \BoxTy{A}
  }[\lockCtx \notin \Gamma']{%
    \ProdCtx{\ExtCtx{\Gamma}{\lockCtx}}{\Gamma'} \vdashNe \unboxIKTm{t} : A
  }\label{rule:unboxNf/IKC-0}%
\end{mathpar}
The normal forms for \ITC, \IKFourC, and \ISFourC are defined similarly
by varying the elimination rule as in their term
typing rules in \cref{fig:it-ik4-is4-syn}.

Following the work on \NbE for \STLC with possible\hyp{}world\footnote{also called ``Kripke'' or ``Kripke\hyp{}style''} models~\parencite{Coquand02},
we instantiate the parameters
that define possible\hyp{}world models for Fitch\hyp{}style calculi as follows: we pick contexts
for $W$, \emph{order\hyp{}preserving embeddings} (sometimes called
``weakenings'', defined in the next \lcnamecref{sec:nbe-models}) $\Gamma \leq \Gamma'$
for $\Gamma \Ri \Gamma'$, and neutral derivations~$\Gamma \vdashNe
\BaseTy$ as the valuation~$V_{\BaseTy,\Gamma}$.
It remains for us to instantiate the parameter $\Rm$ and show that this
model supports the $\quote$ function.

The instantiation of the modal parameter~$\Rm$ in the possible\hyp{}world semantics
varies for each calculus and captures the difference between them.
Recall that the syntaxes of the four calculi only differ in their elimination
rules for $\BoxTy$ types.
When viewed through the lens of the possible\hyp{}world semantics,
this difference can be generalized as follows:
\begin{mathpar}
  \inferrule[$\BoxTy$\nbhyp{}Elim]{%
    \Delta  \vdash t : \BoxTy{A}
  }[(\R{\Delta}{\Gamma})]{%
    \Gamma \vdash \unboxTm{t} : A
  }%
\end{mathpar}
We generalize the relationship between the context in the premise and
the context in the conclusion using a generic modal accessibility
relation $\R{}{}$ between contexts.
When viewed as a candidate for instantiating the $\Rm$ relation,
this rule states that if $\BoxTy{A}$ is derivable in some past world~$\Delta$,
then we may derive $A$ in the current world~$\Gamma$.
The various $\BoxTy$\nbhyp{}elimination rules for Fitch\hyp{}style calculi
can be viewed as instances of this generalized rule, where we define
$\R{}{}$ in accordance with $\BoxTy$\nbhyp{}elimination rule of the calculus
under consideration.
For example, for \IKC, we observe that the context of the premise in
\cref{rule:unboxTm/IKC} is $\Gamma$ and that of the conclusion
is $\ProdCtx{\ExtCtx{\Gamma}{\lockCtx}}{\Gamma'}$ such that
$\lockCtx \not\in \Gamma'$, and thus define $\RIKC{\Delta}{\Gamma}$
as $\exists \Delta'.\, \lockCtx \not\in \Delta' \land \Gamma = \ProdCtx{\ExtCtx{\Delta}{\lockCtx}}{\Delta'}$.
Similarly, we define $\RISFourC{\Delta}{\Gamma}$ as
$\exists \Delta'.\, \Gamma = \ProdCtx{\Delta}{\Delta'}$ for \ISFourC,
and follow this recipe for \ITC and \IKFourC.
Accordingly, we instantiate the $\Rm$ parameter in the \NbE model
with the corresponding definition of $\R{}{}$ in the calculus under consideration.

A key component of implementing the $\quote$ function in \NbE models
is \emph{reification}, which is implemented by a family of
functions~$\reify_{A} : \forall \Gamma .\, \eval{A}_\Gamma \to \Gamma \vdashNf A$
indexed by a type~$A$.
While its implementation for the simply\hyp{}typed fragment follows the standard,
for the modal fragment we are required to give an implementation of
$\reify_{\BoxTy{A}} : \forall \Gamma .\, \eval{\BoxTy{A}}_\Gamma \to \Gamma \vdashNf \BoxTy{A}$.
To reify a value of $\eval{\BoxTy{A}}_\Gamma$, we first observe that
$\eval{\BoxTy{A}}_\Gamma = \forall \Gamma'.\, \Wk{\Gamma}{\Gamma'} \to \forall \Delta.\, \R{\Gamma'}{\Delta} \to \eval{A}_{\Delta}$
by definition of $\eval$ and the instantiations of $\Ri$ with $\Wk{}{}$ and $\Rm$ with $\R{}{}$.
By picking $\Gamma$ for $\Gamma'$ and $\ExtCtx{\Gamma}{\lockCtx}$ for $\Delta$, we get $\eval{A}_{\ExtCtx{\Gamma}{\lockCtx}}$
since $\Wk{}{}$ is reflexive and it can be shown that $\R{\Gamma}{\ExtCtx{\Gamma}{\lockCtx}}$ holds
for the calculi under consideration.
By reifying the value~$\eval{A}_{\ExtCtx{\Gamma}{\lockCtx}}$ recursively, we get a normal
form~$\ExtCtx{\Gamma}{\lockCtx} \vdashNf n : A$, which can be used to construct
the desired normal form~$\Gamma \vdashNf \boxTm{n} : \BoxTy{A}$ using
the rule~\labelcref{rule:boxNf}.

\section{Possible\texorpdfstring{\hyp{}}{-}World Semantics and \texorpdfstring{\NbE}{NbE}}\label{sec:nbe-models}

In this \lcnamecref{sec:nbe-models}, we elaborate on the previous
\lcnamecref{sec:main-idea} by defining possible\hyp{}world models and showing that
Fitch\hyp{}style calculi can be interpreted soundly in these models.
Following this, we outline the details of constructing \NbE models as instances.
We begin with the calculus~\IKC, and then show how the same results can be
achieved for the other calculi.

Before discussing a concrete calculus, we present some of their
commonalities.

\paragraph{Types, Contexts and Order\texorpdfstring{\hyp{}}{-}Preserving Embeddings}

The grammar of types and typing contexts for Fitch\hyp{}style is the
following.
\begin{align*}
  \Ty\quad A \Coloneqq \BaseTy\ |\ A \FunTy B\ |\ \BoxTy{A} &  &\Ctx\quad \Gamma \Coloneqq \EmptyCtx\ |\ \ExtCtx{\Gamma}{A}\ |\ \ExtCtx{\Gamma}{\lockCtx}
\end{align*}
Types are generated by an uninterpreted base type~$\BaseTy$, function types~$A
\FunTy B$, and modal types~$\BoxTy{A}$, and typing contexts are ``snoc''
lists of types and locks.

We define the relation of \emph{order\hyp{}preserving
embeddings}~(\OPE) on typing contexts in \cref{fig:ope}.
An \OPE~$\Wk{\Gamma}{\Gamma'}$ embeds the context~$\Gamma$ into
another context~$\Gamma'$ while preserving the order of types and the
order and number of locks in $\Gamma$.

\begin{figure}[ht]
  \begin{mathpar}
    \inferrule[]{%
    }{%
      \baseWk{} : \Wk{\EmptyCtx}{\EmptyCtx}
    }\label{rule:baseWk}%

    \inferrule[]{%
      o : \Wk{\Gamma}{\Gamma'}
    }{%
      \dropWk{o} : \Wk{\Gamma}{\ExtCtx{\Gamma'}{A}}
    }\label{rule:dropWk}%

    \inferrule[]{%
      o : \Wk{\Gamma}{\Gamma'}
    }{%
      \keepVarWk{o} :\Wk{\ExtCtx{\Gamma}{A}}{\ExtCtx{\Gamma'}{A}}
    }\label{rule:keepVarWk}%

    \inferrule[]{%
      o : \Wk{\Gamma}{\Gamma'}
    }{%
      \keepLockWk{o} : \Wk{\ExtCtx{\Gamma}{\lockCtx}}{\ExtCtx{\Gamma'}{\lockCtx}}
    }\label{rule:keepLockWk}%
  \end{mathpar}
  \caption{Order\hyp{}preserving embeddings}
  \label{fig:ope}
\end{figure}

\subsection{The Calculus~\texorpdfstring{\IKC}{IKC}}\label{sec:ikc}

\subsubsection{Terms, Substitutions and Equational Theory}

To define the intrinsically\hyp{}typed syntax and equational theory
of \IKC, we first define a modal accessibility relation on
contexts~$\RIKC{\Delta}{\Gamma}$, which expresses that context
$\Gamma$ extends $\ExtCtx{\Delta}{\lockCtx}$ to the right without
adding locks.
Note that $\RIKC{\Delta}{\Gamma}$ exactly when $\exists \Delta'.\, \lockCtx
\not\in \Delta' \land \Gamma = \ExtCtx{\ExtCtx{\Delta}{\lockCtx}}{\Delta'}$.
\begin{figure}[ht]
  \begin{mathpar}
    \inferrule[]{%
    }{%
      \nilExt : \RIKC{\Gamma}{\ExtCtx{\Gamma}{\lockCtx}}
    }\label{rule:emptyExt}%

    \inferrule[]{%
      e : \RIKC{\Delta}{\Gamma}
    }{%
      \varExt{e} : \RIKC{\Delta}{\ExtCtx{\Gamma}{A}}
    }\label{rule:varExt}%
  \end{mathpar}
  \caption{Modal accessibility relation on contexts (\IKC)}
  \label{fig:ikc-ce}
\end{figure}

\begin{figure}[ht]
  \begin{mathpar}
    \inferrule[Var\nbhyp{}Zero]{%
    }{%
      \ExtCtx{\Gamma}{A} \vdashVar \zeroVar : A
    }\label{rule:zeroVar}%

    \inferrule[Var\nbhyp{}Succ]{%
      \Gamma \vdashVar v : A
    }{%
      \ExtCtx{\Gamma}{B} \vdashVar \succVar{v} : A
    }\label{rule:succVar}%

    \inferrule[Var]{%
      \Gamma \vdashVar v : A
    }{%
      \Gamma \vdash \varTm{v} : A
    }\label{rule:varTm}%

    \inferrule[$\FunTy$\nbhyp{}Intro]{
      \ExtCtx{\Gamma}{A} \vdash t : B
    }{%
      \Gamma \vdash \lamTm{t} : A \FunTy B
    }\label{rule:lamTm}%

    \inferrule[$\FunTy$\nbhyp{}Elim]{%
      \Gamma \vdash t : A \FunTy B\\
      \Gamma \vdash u : A
    }{%
      \Gamma \vdash \appTm{t,u} : B
    }\label{rule:appTm}%

    \inferrule[$\BoxTy$\nbhyp{}Intro]{%
      \ExtCtx{\Gamma}{\lockCtx} \vdash t : A
    }{%
      \Gamma \vdash \boxTm{t} : \BoxTy{A}
    }\label{rule:boxTm}%

    \inferrule[\IKC/$\BoxTy$\nbhyp{}Elim]{%
      \Delta  \vdash t : \BoxTy{A}\\
      e : \RIKC{\Delta}{\Gamma}
    }{%
      \Gamma \vdash \unboxIKTm{t,e} : A
    }\label{rule:unboxTm/IKC}%
  \end{mathpar}
  \caption{Intrinsically\hyp{}typed terms of \IKC}
  \label{fig:ikc-syn-full}
\end{figure}

\Cref{fig:ikc-syn-full} presents the intrinsically\hyp{}typed syntax of
\IKC.
We will use both $\Gamma \vdash t : A$ and $t : \Gamma \vdash A$ to
say that $t$ denotes an (intrinsically\hyp{}typed) term of type~$A$
in context~$\Gamma$, and similarly for substitutions, which will be
defined below.
Instead of named variables as in \cref{fig:ikc-syn}, variables are
defined using De~Bruijn indices in a separate judgement~$\Gamma
\vdashVar A$ .
The introduction and elimination rules for function types are like
those in \STLC, and the introduction rule for the type~$\BoxTy{A}$ is
similar to that of \cref{fig:ikc-syn}.
The elimination rule~\labelcref{rule:unboxTm/IKC-0} is defined using the
modal accessibility relation~$\RIKC{\Delta}{\Gamma}$ which relates the
contexts in the premise and the conclusion, respectively.
This relation replaces the side condition~$(\lockCtx \not\in
\Gamma')$ in \cref{fig:ikc-syn} and other $\BoxTy$\nbhyp{}elimination
rules in \cref{sec:introduction,sec:main-idea}.
Note that formulating the rule for the term~$\unboxIKTm$ with~$e
:\RIKC{\Delta}{\Gamma}$ as a second premise is in sharp contrast to
\textcite[Fig.~1]{Clouston18} where the relation is not mentioned in
the term but formulated as the \emph{side condition}~$\Gamma =
\ProdCtx{\ExtCtx{\Delta}{\lockCtx}}{\Gamma'}$ for some
lock\hyp{}free~$\Gamma'$.

A term~$\Gamma \vdash t : A$ can be \emph{weakened}, which is a
special case of \emph{renaming}, with an \OPE (see \cref{fig:ope}) using a function~$\wk :
\Wk{\Gamma}{\Gamma'} \to \Gamma \vdash A \to \Gamma' \vdash A$.
Given an \OPE~$o : \Wk{\Gamma}{\Gamma'}$, renaming the term using
$\wk$ yields a term~$\Gamma' \vdash \wk{o,t} : A$ in the weaker
context~$\Gamma'$.
The unit element for $\wk$ is the identity \OPE~$\idWk :
\Wk{\Gamma}{\Gamma}$, \ie $\wk{\idWk,t} = t$.
Renaming arises naturally when evaluating terms and in specifying the
equational theory (\eg in the $\eta$ rule of function type).

\begin{figure}[ht]
  \begin{mathpar}
    \inferrule[]{%
    }{%
      \Gamma \vdashSub \emptySub : \EmptyCtx
    }\label{rule:emptySub}

    \inferrule[]{%
      \Gamma \vdashSub s : \Delta\\
      \Gamma \vdash t : A
    }{%
      \Gamma \vdashSub \extTmSub{s,t} : \ExtCtx{\Delta}{A}
    }\label{rule:extTmSub}

    \inferrule[]{%
      \Theta \vdashSub s : \Delta\\
      e : \RIKC{\Theta}{\Gamma}
    }{%
      \Gamma \vdashSub \extLockSub{s,e} : \ExtCtx{\Delta}{\lockCtx}
    }\label{rule:extLockSub/IKC}
  \end{mathpar}
  \caption{Substitutions for \IKC}
  \label{fig:ikc-sub}
\end{figure}

Substitutions for \IKC are inductively defined in \cref{fig:ikc-sub}.
A judgement~$\Gamma \vdashSub s : \Delta$ denotes a substitution for a
context~$\Delta$ in the context~$\Gamma$.
Applying a substitution to a term~$\Delta \vdash t : A$, \ie
$\subst{s,t} : \Gamma \vdash A$, yields a term in the context
$\Gamma$.
The substitution~$\idSub : \Gamma \vdashSub \Gamma$ denotes the
identity substitution, which exists for all $\Gamma$.
As usual, it can be shown that terms are closed under the application
of a substitution, and that it preserves the identity, \ie $\subst{\idSub,t} = t$.
Substitutions are also closed under renaming and this operation
preserves the identity as well.

The equational theory for \IKC, omitting congruence rules,
is specified in \cref{fig:ikc-conversion-full}.
As discussed earlier, \IKC extends the usual rules in \STLC
(\cref{ik:rule:fun-beta,ik:rule:fun-eta}) with rules for the $\BoxTy$
type (\cref{ik:rule:box-beta,ik:rule:box-eta}).
The function~$\factor : \RIKC{\Delta}{\Gamma} \to
\Wk{\ExtCtx{\Delta}{\lockCtx}}{\Gamma}$, in \cref{ik:rule:box-beta},
maps an element of the modal accessibility relation~$e :
\RIKC{\Delta}{\Gamma}$ to an
OPE~$\Wk{\ExtCtx{\Delta}{\lockCtx}}{\Gamma}$.
This is possible because the context~$\Gamma$ does not have any lock
to the right of $\ExtCtx{\Delta}{\lockCtx}$.
\begin{figure}[ht]
  \begin{mathpar}
    \inferrule[$\FunTy$\nbhyp{}$\beta$]{ %
      \ExtCtx{\Gamma}{A} \vdash t : B\\
      \Gamma \vdash u : A
    }{%
      \Gamma \vdash \appTm{(\lamTm{t}),u} \thyeq \subst{(\extTmSub{\idSub,u}),t}
    }\label{ik:rule:fun-beta}%

    \inferrule[$\FunTy$\nbhyp{}$\eta$]{ %
      \Gamma \vdash t : A \FunTy B\\
    }{%
    \Gamma \vdash t \thyeq \lamTm{(\appTm{(\wk{(\dropIdWk),t}),(\varTm{\zeroVar})})}
    }\label{ik:rule:fun-eta}%

    \inferrule[$\BoxTy$\nbhyp{}$\beta$]{ %
      \ExtCtx{\Delta}{\lockCtx} \vdash t : A\\
      e : \RIKC{\Delta}{\Gamma}
    }{%
      \Gamma \vdash \unboxIKTm{(\boxTm{t}),e} \thyeq \wk{(\factor{e}),t}
    }\label{ik:rule:box-beta}%

    \inferrule[$\BoxTy$\nbhyp{}$\eta$]{ %
      \Gamma \vdash t : \BoxTy{A}\\
    }{%
      \Gamma \vdash t \thyeq \boxTm{(\unboxIKTm{t, \nilExt})}
    }\label{ik:rule:box-eta}%
  \end{mathpar}
  \caption{Equational theory for \IKC}
  \label{fig:ikc-conversion-full}
\end{figure}

\subsubsection{Possible\texorpdfstring{\hyp{}}{-}World Semantics}

A possible\hyp{}world model is defined using the notion of a
possible\hyp{}world frame as below.
We work in a constructive type\hyp{}theoretic metalanguage,
and denote the universe of types in this language by $\Type$.
\begin{definition}[Possible\hyp{}world frame]\label{def:frame}
  A frame~$F$ is given by a triple~$\triple{W}{\Ri}{\Rm}$ consisting of
  a type~$W : \Type$ and two relations~$\Ri$ and~$\Rm : W \times W \to \Type$ on $W$ such that
  the following conditions are satisfied:
  \begin{itemize}
    \item $\Ri$ is \emph{reflexive} and \emph{transitive}
    \item if $w \Rm v$ and $v \Ri v'$ then there exists some $w' : W$
      such that $w \Ri w'$ and $w' \Rm v'$; this \emph{factorization} condition can be
      pictured as an
      implication~${\Rm} \semi {\Ri} \subseteq {\Ri} \semi {\Rm}$ or
      diagrammatically as follows:
      
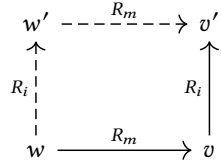
\begin{center}
        \begin{tikzcd}[scale=2, sep=huge]
          w' \arrow[r, "\Rm", dashrightarrow]                   & v'\\
          w  \arrow[u, "\Ri", dashrightarrow] \arrow[r, "\Rm"]  & v \arrow[u, "\Ri"]
        \end{tikzcd}
      \end{center}
      (note that neither $w'$ nor the proofs of relatedness are
      required to be unique, nor will they all be in the frames that we
      will consider)
  \end{itemize}
\end{definition}
\begin{definition}[Possible\hyp{}world model]\label{def:possible-world-model}
  A possible\hyp{}world model~$\Mod$ is given by a tuple~$\tuple{F,V}$
  consisting of a frame~$F$ (see \cref{def:frame}) and a $W$\nbhyp{}indexed family~$V_\BaseTy : W \to \Type$ (called the \emph{valuation} of the base type)
    such that
    $\forall w, w'.\, w \Ri w' \to
    V_{\BaseTy,w} \to V_{\BaseTy,w'}$.
\end{definition}
We have omitted coherence conditions from these
\lcnamecrefs{def:frame} for readability.
Those conditions stem from the proof relevance of the relations
and predicates
involved.
They will be satisfied by the models we will construct, and will also
be given below for completeness.

The types and typing contexts in \IKC are interpreted in a
possible\hyp{}world model via the interpretation functions~$\eval$
defined in \cref{para:possible-world}.
To evaluate terms, we must first prove the following \emph{monotonicity} lemma.
This lemma is well\hyp{}known as a requirement to give
a sound interpretation of the function type in an
arbitrary possible\hyp{}world model, and can be thought
of as the semantic generalization of renaming in terms.
\begin{lemma}[Monotonicity]
  In every possible\hyp{}world model~$\Mod$, for every type~$A$ and worlds $w$ and $w'$,
  we have a function $\wk_A : w \Ri w' \to \eval{A}_w \to \eval{A}_{w'}$.
  And similarly, for every context $\Gamma$, a function~$\wk_\Gamma : w \Ri w' \to \eval{\Gamma}_w \to \eval{\Gamma}_{w'}$.
\end{lemma}

We evaluate terms in \IKC in a possible\hyp{}world model as follows.
\begin{equation*}
  \begin{array}{>{\evallbracket}l@{\evalrbracket} @{\$} l @{\;}c@{\;} l}
    \multicolumn{4}{l}{\eval : \Gamma \vdash A \to (\forall w.\, \eval{\Gamma}_{w} \to  \eval{A}_{w})} \\
    \varTm{v}       & \gamma & = & \lookupTmS{v,\gamma} \\
    \lamTm{t}       & \gamma & = & \lambda i.\, \lambda a.\, \eval{t}\$\pair{\wk{i,\gamma}}{a} \\
    \appTm{t,u}     & \gamma & = & (\eval{t}\$ \gamma)\$ \idWk\$ (\eval{u}\$ \gamma) \\
    \boxTm{t}       & \gamma & = & \lambda i.\, \lambda m.\, \eval{t}\$\pair{\wk{i,\gamma}}{m} \\
    \unboxIKTm{t,e} & \gamma & = & \eval{t}\$ \delta\$ \idWk\$ m \\
    \multicolumn{4}{l}{\quad\quad \text{where } \pair{\delta}{m} = \trimIKSubS{\gamma,e}}
  \end{array}
\end{equation*}

The evaluation of terms in the simply\hyp{}typed fragment is standard,
and resembles the evaluator of \STLC.
Variables are interpreted by a lookup function that projects values from
an environment, and $\lambda$\nbhyp{}abstraction and application are
evaluated using their semantic counterparts.
To evaluate $\lambda$\nbhyp{}abstraction, we must construct a semantic function~$\forall w' .\, w \Ri w' \to \eval{A}_{w'} \to \eval{B}_{w'}$
using the given term  $\ExtCtx{\Gamma}{A} \vdash t : B$ and
environment $\gamma : \eval{\Gamma}_w$.
We achieve this by recursively evaluating $t$ in an environment that
extends $\gamma$ appropriately using the semantic arguments
$i : w \Ri w'$ and $a : \eval{A}_{w'}$.
We use the monotonicity lemma to ``transport''
$\eval{\Gamma}_w$ to $\eval{\Gamma}_{w'}$, and construct
an environment of type $\eval{\Gamma}_{w'} \times \eval{A}_{w'}$
for recursively evaluating $t$, which produces the desired result
of type $\eval{B}_{w'}$.
Application is evaluated by simply recursively evaluating
the applied terms and applying them in the semantics
with a value~$\idWk : w \Ri w$, which is available
since $\Ri$ is reflexive.

In the modal fragment, to evaluate the term~$\Gamma \vdash \boxTm{t} : \BoxTy{A}$
with $\gamma : \eval{\Gamma}_w$, we must construct a function of
type $\forall w'.\, w \Ri w' \to \forall v.\, w' \Rm v \to \eval{A}_{v}$.
Using the semantic arguments~$i : w \Ri w'$ and $m : w' \Rm v$, we recursively evaluate
the term $\ExtCtx{\Gamma}{\lockCtx} \vdash t : A $ in the
extended environment~$\pair{\wk{i,\gamma}}{m} : \eval{\ExtCtx{\Gamma}{\lockCtx}}_v$,
since $\eval{\ExtCtx{\Gamma}{\lockCtx}}_v = \textstyle\sum_{w'} {\eval{\Gamma}_{w'} \times w' \Rm v}$.
On the other hand, the term~$\Gamma \vdash \unboxIKTm{t,e} : A$
with $e : \RIKC{\Delta}{\Gamma}$ and
$\Delta \vdash t : \BoxTy{A}$, for some $\Delta$,
must be evaluated with an environment~$\gamma : \eval{\Gamma}_w$.
To recursively evaluate the term~$\Delta \vdash t : \BoxTy{A}$, we must first
discard the part of the environment~$\gamma$ that substitutes the types
in the extension of $\ExtCtx{\Delta}{\lockCtx}$.
This is achieved using the
function~$\trimIKSubS : \eval{\Gamma}_w \to  \RIKC{\Delta}{\Gamma} \to \eval{\ExtCtx{\Delta}{\lockCtx}}_w$
that projects $\gamma$ to produce
an environment~$\delta : \eval{\Delta}_{v'}$ and a value~$m : v' \Rm w$.
We evaluate $t$ with $\delta$ and apply
the resulting function of type~$\forall v.\, v \Ri v' \to \forall w.\, v' \Rm w \to \eval{A}_{w}$
to $\idWk$ and $m$ to return the desired result.

We state the soundness of \IKC with respect to the possible\hyp{}world
semantics before we instantiate it with the \NbE model that we will
construct in the next \lcnamecref{sec:ikc-nbe-model}.
We note that the soundness proof relies on the possible\hyp{}world
models to satisfy coherence conditions that we have omitted from
\cref{def:frame,def:possible-world-model} but that will be satisfied
by the \NbE models. Specifically, $W$ and $\Ri$ together with the
transitivity and reflexivity proofs~$\transRi$ and~$\reflRi$ for $\Ri$
need to form a category~$\mathscr{W}$, \ie $\transRi$ needs to be
associative and $\reflRi$ needs to be a unit for $\transRi$; the
proofs of the factorization condition need to satisfy the
functoriality
laws~$\factorRi{m,(\reflRi{v})} =
\reflRi{w}$,~$\factorRm{m,(\reflRi{v})} =
m$,~$\factorRi{m,(\transRi{i,j})} =
\transRi{(\factorRi{m,i}),(\factorRi{m',j})}$
and~$\factorRm{m,(\transRi{i,j})} = \factorRm{m',j}$ where
$m' \coloneqq \factorRm{m,i} : w' \Rm v'$ denotes the modal
accessibility proof produced by the first factorization of
$m : w \Rm v$ and $i : v \Ri v'$; and $V_{\BaseTy}$ together with the
monotonicity proof~$\wk_{\BaseTy}$ needs to form a functor on the
category~$\mathscr{W}$, \ie $\wk_{\BaseTy}{(\reflRi{w})}$ needs to be
equal to the identity function on $V_{\BaseTy,w}$ and
$\wk_{\BaseTy}{(\transRi{i,j})}$ needs to be equal to the composite
$\wk_{\BaseTy}{j} \after \wk_{\BaseTy}{i}$.

\begin{theorem}\label{thm:ikc-sound-possible-world}
  Let $\Mod$ be any possible\hyp{}world
  model (see \cref{def:possible-world-model}). If two terms~$t$
  and~$u : \Tm{\Gamma}{A}$ of \IKC are
  equivalent (see \cref{fig:ikc-conversion-full})
  then the functions~$\eval{t}$ and~$\eval{u} : \forall w.\, \eval{\Gamma}_{w} \to \eval{A}_{w}$ as determined by $\Mod$ are
  equal.

  \begin{proof}
    \def\C{\Cat[C]}
    Let $\Mod$ be a possible\hyp{}world model with underlying
    frame~$F = \triple{W}{\Ri}{\Rm}$.
    Denote the category whose objects are worlds~$w : W$ and whose
    morphisms are proofs~$i : w \Ri w'$ by $\C$.
    The frame~$F$ can be seen as determining an
    adjunction~$\lockCtx \dashv \BoxTy$ on the category of presheaves
    indexed by the category~$\C$, which is moreover well\hyp{}known to
    be Cartesian closed.
    The interpretation~$\eval$
    can then be seen as factoring through the categorical semantics
    described in \textcite[Section~2.3]{Clouston18}, of which the
    category of presheaves over $\C$ is an instance by virtue of its
    Cartesian closure and equipment with an adjunction.
    We can therefore conclude by applying
    \textcite[Theorem~2.8~(Categorical Soundness) and remark below
    that]{Clouston18}.
  \end{proof}
\end{theorem}

\subsubsection{\texorpdfstring{\NbE}{NbE} Model}\label{sec:ikc-nbe-model}

The normal forms of terms in \IKC are defined along with neutral elements
in a mutually recursive fashion by the judgements~$\Gamma \vdashNf A$ and~$\Gamma \vdashNe A$, respectively, in \cref{fig:ikc-nf}.
Intuitively, a normal form may be thought of as a value, and
a neutral element may be thought of as a ``stuck'' computation.
We extend the standard definition of normal forms and neutral elements in \STLC
with \cref{rule:boxNf,rule:unboxNe/IKC}.

\begin{figure}[ht]
  \begin{mathpar}
    \inferrule[Ne/Var]{%
      \Gamma \vdashVar v : A
    }{%
      \Gamma \vdashNe \varTm{v} : A
    }\label{rule:varNe}%

    \inferrule[Nf/Up]{%
      \Gamma \vdashNe n : \BaseTy
    }{%
      \Gamma \vdashNf \upTm{n} : \BaseTy
    }\label{rule:upNf}%

    \inferrule[Nf/$\FunTy$\nbhyp{}Intro]{%
      \ExtCtx{\Gamma}{A} \vdashNf n : B
    }{%
      \Gamma \vdashNf \lamTm{n} : A \FunTy B
    }\label{rule:lamNf}%

    \inferrule[Ne/$\FunTy$\nbhyp{}Elim]{%
      \Gamma \vdashNe n : A \FunTy B\\
      \Gamma \vdashNf m : A
    }{%
      \Gamma \vdashNe \appTm{n,m} : B
    }\label{rule:appNe}%

    \inferrule[Nf/$\BoxTy$\nbhyp{}Intro]{
      \ExtCtx{\Gamma}{\lockCtx} \vdashNf n : A
    }{
      \Gamma \vdashNf \boxTm{n} : \BoxTy{A}
    }\label{rule:boxNf}

    \inferrule[\IKC/Ne/$\BoxTy$\nbhyp{}Elim]{%
      \Delta  \vdashNe n : \BoxTy{A}\\
      e : \RIKC{\Delta}{\Gamma}
    }{%
      \Gamma \vdashNe \unboxIKTm{n,e} : A
    }\label{rule:unboxNe/IKC}%
  \end{mathpar}
  \caption{Normal forms and neutral elements in \IKC}
  \label{fig:ikc-nf}
\end{figure}

\def\TT{\systemfont C\xspace}
Recall that an \NbE model for a given calculus~\TT is a particular
kind of model~$\Mod$ that comes equipped with a
function~$\quote : \Mod(\eval{\Gamma},\eval{A}) \to \Gamma \vdashNf A$
satisfying~$t \thyeq \quote{\eval{t}}$ for all
terms~$t : \Tm{\Gamma}{A}$ where $\eval$ denotes the \emph{generic}
evaluation function for \TT.

Using the relations defined in \cref{fig:ope,fig:ikc-ce}, we construct an \NbE model for \IKC by
instantiating the parameters that define a \emph{possible\hyp{}world} model as follows.
\begin{itemize}
  \item Worlds as contexts: $W = \Ctx$
  \item Relation~$\Ri$ as order\hyp{}preserving embeddings: $\Gamma \Ri \Gamma' = \Wk{\Gamma}{\Gamma'}$
  \item Relation~$\Rm$ as extensions of a ``locked'' context: $\Delta \Rm \Gamma = \RIKC{\Delta}{\Gamma}$
  \item Valuation $V_\BaseTy$ as neutral elements: $V_{\BaseTy,\Gamma} = \Gamma \vdashNe \BaseTy$
\end{itemize}
The condition that the valuation must satisfy
$\wk_A : \Wk{\Gamma}{\Gamma'} \to \Gamma \vdashNe A \to \Gamma' \vdashNe A$,
for all types $A$, can be shown by induction on the neutral term~$\Gamma \vdashNe A$.
To show that this model is indeed a possible\hyp{}world model, it
remains for us to show that the frame conditions are satisfied.

The first frame condition states that \OPEs must be
reflexive and transitive, which can be shown by structural
induction on the context and definition of \OPEs,
respectively.
The second frame condition states that given%
~$\RIKC{\Delta}{\Gamma}$ and $\Wk{\Gamma}{\Gamma'}$
there is a~$\Delta' : \Ctx$ such that $\Wk{\Delta}{\Delta'}$ and~$\RIKC{\Delta'}{\Gamma'}$,
\begin{center}
  \begin{tikzcd}[scale=2, sep=huge]
    \Delta' \arrow[r, "\RIKC{}{}", dashrightarrow]                       & \Gamma'\\
    \Delta  \arrow[u, "\Wk{}{}", dashrightarrow] \arrow[r, "\RIKC{}{}"]  & \Gamma \arrow[u, "\Wk{}{}"]
  \end{tikzcd}
\end{center}
which can be shown by constructing a
function
by simultaneous recursion on \OPEs and the modal accessibility
relation.

Observe that the instantiation of the monotonicity lemma in the \NbE model states that we
have the functions~$\wk_A : \Wk{\Gamma}{\Gamma'} \to \eval{A}_\Gamma \to \eval{A}_{\Gamma'}$
and~$\wk_\Delta : \Wk{\Gamma}{\Gamma'} \to \eval{\Delta}_\Gamma \to \eval{\Delta}_{\Gamma'}$,
which allow denotations of types and contexts to be renamed with respect to an \OPE.

To implement the function $\quote$, we first implement \emph{reification} and \emph{reflection},
using two functions~$\reify_{A} : \eval{A}_{\Gamma} \to \Gamma \vdashNf A$
and~$\reflect_{A} : \Gamma \vdashNe A \to \eval{A}_{\Gamma}$, respectively.
Reification converts a semantic value to a normal form, while reflection
converts a neutral element to a semantic value.
They are implemented as follows by induction on the index type $A$.
\begin{align*}
  & \begin{array}{l @{\$} l @{\;}c@{\;} l}
      \multicolumn{4}{l}{\reify_{A,\Gamma} : \eval{A}_{\Gamma} \to \Gamma \vdashNf A} \\
      \reify_{\BaseTy,\Gamma}    & n & = & \upTm{n} \\
      \reify_{A \FunTy B,\Gamma} & f & = & \lamTm{(\reify_{B,(\ExtCtx{\Gamma}{A})}(f\$ (\dropIdWk)\$ \fresh_{A,\Gamma}))} \\
      \reify_{\BoxTy{A},\Gamma}  & b & = & \boxTm{(\reify_{A,(\ExtCtx{\Gamma}{\lockCtx})}(b\$ \idWk\$ \nilExt))}
    \end{array}\\[\the\abovedisplayskip]
  & \begin{array}{l @{\$} l @{\;}c@{\;} l}
      \multicolumn{4}{l}{\reflect_{A,\Gamma} : \Gamma \vdashNe A \to \eval{A}_{\Gamma}} \\
      \reflect_{\BaseTy,\Gamma}    & n & = & n \\
      \reflect_{A \FunTy B,\Gamma} & n & = & \lambda (o : \Wk{\Gamma}{\Gamma'}).\, \lambda a .\, {\reflect_{B,\Gamma}(\appTm{(\wk_{A \FunTy B}{o,n}),(\reify_{A,\Gamma'}{a}))}} \\
      \reflect_{\BoxTy{A},\Gamma}  & n & = & \lambda (o : \Wk{\Gamma}{\Gamma'}).\, \lambda (e : \RIKC{\Gamma'}{\Delta}).\, \reflect_{A,\Delta}(\unboxIKTm{(\wk_{\BoxTy{A}}{o,n}),e})
    \end{array}
\end{align*}

For the function type, we recursively reify the body of the $\lambda$\nbhyp{}abstraction
by applying the given semantic function~$f$ with suitable arguments, which are
an \OPE $\dropIdWk : \Wk{\Gamma}{\ExtCtx{\Gamma}{A}}$
and a value~$\fresh_{A,\Gamma} = \reflect_{A,(\ExtCtx{\Gamma}{A})}{(\varTm{\zeroVar})} : \eval{A}_{\ExtCtx{\Gamma}{A}}$\mdash{}which
is the De Bruijn index equivalent of a fresh variable.
Reflection, on the other hand, recursively reflects the application of a
neutral~$\Gamma \vdashNe n : A \FunTy B$
to the reification of the semantic argument~$a : \eval{A}_{\Gamma'}$
for an \OPE~$o : \Wk{\Gamma}{\Gamma'}$.
Similarly, for the $\BoxTy$ type, we recursively reify the body of
$\boxTm$ by applying the given semantic
function~$b : \forall \Gamma.\, \Wk{\Gamma}{\Gamma'} \to \forall \Delta .\, \RIKC{\Gamma'}{\Delta} \to \eval{A}_\Delta$
to suitable arguments $\idWk : \Wk{\Gamma}{\Gamma}$ and
the empty context extension $\nilExt : \RIKC{\Gamma}{\ExtCtx{\Gamma}{\lockCtx}}$.
Reflection also follows a similar pursuit by reflecting the application of the
neutral~$\Gamma \vdashNe n : \BoxTy{A}$ to the eliminator $\unboxTm$.

Equipped with reification, we implement $\quote$ (as seen below),
by applying the given denotation of a term,
a function~$f : \forall \Delta .\, \eval{\Gamma}_{\Delta} \to  \eval{A}_{\Delta}$,
to the identity environment $\freshEnv_{\Gamma} : \eval{\Gamma}_\Gamma$,
and then reifying the resulting value.
The construction of the value~$\freshEnv_{\Gamma}$ is the De Bruijn index equivalent
of generating an environment with fresh variables.
\begin{align*}
  & \begin{array}{l @{\;}c@{\;} l}
      \multicolumn{3}{l}{\quote : (\forall \Delta .\, \eval{\Gamma}_{\Delta} \to  \eval{A}_{\Delta}) \to \Gamma \vdashNf A} \\
      \quote{f} & = & \reify_{A,\Gamma}{(f\$ \freshEnv_{\Gamma})}
    \end{array}\\[\the\abovedisplayskip]
  & \begin{array}{l @{\;}c@{\;} l}
      \multicolumn{3}{l}{\freshEnv_{\Gamma} : \eval{\Gamma}_{\Gamma}} \\
      \freshEnv_{\EmptyCtx}                 & = & () \\
      \freshEnv_{\ExtCtx{\Gamma}{A}}        & = & \pair{\wk{(\dropIdWk),\freshEnv_{\Gamma}}}{\fresh_{A,\Gamma}} \\
      \freshEnv_{\ExtCtx{\Gamma}{\lockCtx}} & = & \pair{\freshEnv_{\Gamma}}{\nilExt{}}
    \end{array}
\end{align*}

To prove that the function~$\quote$ is indeed a retraction of evaluation,
we follow the usual logical relations approach.
As seen in \cref{fig:ikc-lr}, we define a relation~$\LR_{A}$ indexed by a type~$A$
that relates a term~$\Gamma \vdash t : A$ to its denotation~$a : \eval{A}_\Gamma$
as $\LR_{A}{t,a}$.
From a proof of $\LR_{A}{t,a}$, it can be shown that $t \thyeq \reify_{A}{a}$.
This relation is extended to contexts as $\LR_{\Delta}$, for some context~$\Delta$,
which relates a substitution~$\Gamma \vdash s : \Delta$ to its
denotation~$\delta : \eval{\Delta}_\Gamma$ as $\LR_{\Delta}{s,\delta}$.
\begin{figure}[ht]
  \begin{align*}
    & \begin{array}{l @{\$} l @{\$} l @{\;}c@{\;} l}
        \multicolumn{5}{l}{\LR_{A,\Gamma} : \Gamma \vdash A \to \eval{A}_{\Gamma} \to \Type} \\
        \LR_{\BaseTy,\Gamma}    & t & n & = & t \thyeq \quote{n} \\
        \LR_{A \FunTy B,\Gamma} & t & f & = & \forall \Gamma', o : \Wk{\Gamma}{\Gamma'}, u, a.\, \LR_{A,\Gamma'}{u,a} \to \LR_{B,\Gamma'}{(\appTm{(\wk{o,t}), u}),(f\$ o\$ a)} \\
        \LR_{\BoxTy{A},\Gamma}  & t & b & = & \forall \Gamma', o : \Wk{\Gamma}{\Gamma'}, e : \RIKC{\Gamma'}{\Delta}.\, \LR_{A,\Delta}{(\unboxIKTm{(\wk{o,t}),e}),(b\$ o \$ e)}
      \end{array}\\[\the\intextsep]
    & \begin{array}{l @{\$} l @{\$} l @{\;}c@{\;} l}
        \multicolumn{5}{l}{\LR_{\Delta,\Gamma} : \Gamma \vdashSub \Delta \to \eval{\Delta}_{\Gamma} \to \Type} \\
        \LR_{\EmptyCtx,\Gamma}                   & \emptySub                                    & ()               & = & \top \\
        \LR_{(\ExtCtx{\Delta}{A}),\Gamma}        & (\extTmSub{s,t})                             & \pair{\delta}{a} & = & \LR_{\Delta,\Gamma}{s,\delta} \times \LR_{A,\Gamma}{t,a} \\
        \LR_{(\ExtCtx{\Delta}{\lockCtx}),\Gamma} & (\extLockSub{s,(e : \RIKC{\Theta}{\Gamma})}) & \pair{\delta}{e} & = & \LR_{\Delta,\Theta}{s,\delta}
      \end{array}
  \end{align*}
  \caption{Logical relations for \IKC}
  \label{fig:ikc-lr}
\end{figure}

For the logical relations, we then prove the so\hyp{}called
fundamental theorem.

\begin{proposition}[Fundamental theorem]\label{pro:fundamental-ikc}
  Given a term~$\Delta \vdash t : A$, a substitution~$\Gamma \vdashSub s : \Delta$
  and a value~$\delta : \eval{\Delta}_\Gamma$,
  if $\LR_{\Delta,\Gamma}{s,\delta}$ then $\LR_{A,\Gamma}{(\subst{s,t}),(\eval{t}\$\delta)}$.
\end{proposition}

We conclude this \lcnamecref{sec:ikc} by stating the normalization
theorem for \IKC.

\Cref{pro:fundamental-ikc} entails that
$\LR_{A,\Delta}{(\subst{\idSub,t}),(\eval{t}\$\freshEnv_{\Delta})}$ for any
term~$t$, if we pick $s$ as the identity
substitution~$\idSub : \Delta \vdashSub \Delta$, and $\delta$ as
$\freshEnv_{\Delta} : \eval{\Delta}_\Delta$, since they can be shown
to be related as $\LR_{\Delta,\Delta}{\idSub,\freshEnv_{\Delta}}$.
From this it follows that $\subst{\idSub,t} \thyeq \reify_{A}{(\eval{t}\$\freshEnv_{\Delta})}$,
and further that $t \thyeq \quote{\eval{t}}$ from the definition of $\quote$
and the fact that $\subst{\idSub,t} = t$.
As a result, the composite~$\norm = \quote \circ \eval$ is
\emph{adequate}%
, \ie $\norm{t} = \norm{t'}$ implies $t \thyeq t'$.

The soundness of \IKC with respect to possible\hyp{}world
models (see \cref{thm:ikc-sound-possible-world}) directly entails
$\quote{\eval{t}} = \quote{\eval{u}} : \Nf{\Gamma}{A}$ for all terms
$t$, $u : \Tm{\Gamma}{A}$ such that $\Gamma \vdash t \thyeq u : A$,
which means that $\norm = \quote \circ \eval$ is \emph{complete}. Note
that this terminology might be slightly confusing
because it is the \emph{soundness} of $\eval$ that implies the
\emph{completeness} of $\norm$.

\begin{theorem}\label{thm:ikc-normalization}
  Let $\Mod$ denote the possible\hyp{}world model over the frame given
  by the relations~$\Wk{\Gamma}{\Gamma'}$ and~$\RIKC{\Delta}{\Gamma}$
  and the valuation~$V_{\BaseTy,\Gamma} =
  \Ne{\Gamma}{\BaseTy}$.

  There is a
  function~$\quote : \Mod(\eval{\Gamma},\eval{A}) \to \Nf{\Gamma}{A}$
  such that the
  composite~$\norm = \quote \circ \eval : \Tm{\Gamma}{A} \to
  \Nf{\Gamma}{A}$ from terms to normal forms of \IKC is complete and
  adequate.
\end{theorem}

\subsection{Extending to the Calculus~\texorpdfstring{\ISFourC}{IS4C}}
\label{sec:is4c}

\subsubsection{Terms, Substitutions and Equational Theory}
To define the intrinsically\hyp{}typed syntax of \ISFourC, we first define
the modal accessibility relation on contexts in \cref{fig:is4-ce}.
\begin{figure}[ht]
  \begin{mathpar}
    \inferrule[]{%
    }{%
      \nilExt : \RISFourC{\Gamma}{\Gamma}
    }%

    \inferrule[]{%
      e : \RISFourC{\Delta}{\Gamma}
    }{%
      \varExt{e} : \RISFourC{\Delta}{\ExtCtx{\Gamma}{A}}
    }%

    \inferrule[]{%
      e : \RISFourC{\Delta}{\Gamma}
    }{%
      \lockExt{e} : \RISFourC{\Delta}{\ExtCtx{\Gamma}{\lockCtx}}
    }\label{rule:lockExt}%
  \end{mathpar}
  \caption{Modal accessibility relation on contexts (\ISFourC)}
  \label{fig:is4-ce}
\end{figure}

If $\RISFourC{\Delta}{\Gamma}$ then $\Gamma$ is an extension of
$\Delta$ with as many locks as needed.
Note that, in contrast to \IKC, the modal accessibility relation is
both reflexive and transitive.
This corresponds to the conditions on the accessibility relation
for the logic~\ISFour.

\Cref{fig:is4-full} presents the changes of \IKC that yield \ISFourC.
The terms are the same as \IKC with the exception of
\cref{rule:unboxTm/IKC} which now includes the modal accessibility
relation for \ISFourC.
Similarly, the substitution rule for contexts with locks now refers
to $\RISFourC{}{}$.

\begin{figure}[ht]
  \begin{mathpar}
    \inferrule[\ISFourC/$\BoxTy$\nbhyp{}Elim]{%
      \Delta \vdash t : \BoxTy{A}\\
      e : \RISFourC{\Delta}{\Gamma}
    }{%
      \Gamma \vdash \unboxISFourTm{t,e} : A
    }\label{rule:unboxTm/ISFourC}%

    \inferrule*{%
      \Theta \vdash s : \Delta\\
      e : \RISFourC{\Theta}{\Gamma}
    }{%
      \Gamma \vdashSub \extLockSub{s,e} : \ExtCtx{\Delta}{\lockCtx}
    }\label{rule:extLockSub/ISFourC}%
  \end{mathpar}
  \caption{Intrinsically\hyp{}typed terms and substitutions of
    \ISFourC (omitting the unchanged rules of
    \cref{fig:ikc-syn-full})}
  \label{fig:is4-full}
\end{figure}

\Cref{fig:is4c-conversion} presents the equational theory of the modal
fragment of \ISFourC.
This is a slightly modified version of
\IKC (\cf \cref{fig:ikc-conversion-full}) that accommodates the changes to
the rule~\labelcref{rule:unboxTm/ISFourC}.
Unlike before, \cref{is4:rule:box-beta} now performs a substitution
to modify the term~$\ExtCtx{\Delta}{\lockCtx} \vdash t : A$ to a term of type
$\Gamma \vdash A$.
Note that the result of such a substitution need not yield the same
term since substitution may change the context extension of some
subterm.
\begin{figure}[ht]
  \begin{mathpar}
    \inferrule[$\BoxTy$\nbhyp{}$\beta$]{ %
      \ExtCtx{\Delta}{\lockCtx} \vdash t : A\\
      e : \RISFourC{\Delta}{\Gamma}
    }{%
      \Gamma \vdash \unboxISFourTm{(\boxTm{t}),e} \thyeq \subst{(\extLockSub{\idSub,e}),t}
    }\label{is4:rule:box-beta}%

    \inferrule[$\BoxTy$\nbhyp{}$\eta$]{ %
      \Gamma \vdash t : \BoxTy{A}
    }{%
      \Gamma \vdash t \thyeq \boxTm{(\unboxISFourTm{t, (\lockExt{\nilExt})})}
    }\label{is4:rule:box-eta}%
  \end{mathpar}
  \caption{Equational theory for \ISFourC (omitting the unchanged
    rules of \cref{fig:ikc-conversion-full})}
  \label{fig:is4c-conversion}
\end{figure}

\subsubsection{Possible\texorpdfstring{\hyp{}}{-}World Semantics}
Giving possible\hyp{}world semantics for \ISFourC requires an
additional frame condition on the relation $R_m$: it must be
reflexive and transitive.
Evaluation proceeds as before, where we use a function~$\trimISFourSubS : \forall w.\, \eval{\Gamma}_w \to \RISFourC{\Delta}{\Gamma} \to \eval{\Delta , \lockCtx}_w$
to manipulate the environment for evaluating
$\unboxISFourTm{t,e}$, as seen below.
\begin{equation*}
\begin{array}{>{\evallbracket}l@{\evalrbracket} @{\$} l @{\;}c@{\;} l}
  \unboxISFourTm{t,e} & \gamma & = \eval{t}\$\delta\$\idWk\$m \\
  \multicolumn{4}{l}{\quad\quad \text{where } \pair{\delta}{m} = \trimISFourSubS{\gamma,e}}
\end{array}
\end{equation*}
The additional frame requirements ensures that the
function~$\trimISFourSubS$ can be implemented.
For example, consider implementing the case of $\trimISFourSubS$ for
some argument of type $\eval{\Gamma}_w$ and the extension
$\nilExt : \RISFourC{\Gamma}{\Gamma}$ that adds zero locks.
The desired result is of type $\eval{\Gamma , \lockCtx}_w$, which is defined
as $\textstyle\sum_{v} {\eval{\Gamma}_v \times v \Rm w}$.
We construct such a result using the argument of $\eval{\Gamma}_w$
by picking $v$ as $w$ itself, and using the reflexivity of $\Rm$
to construct a value of type $w \Rm w$.
Similarly, the transitivity of $\Rm$ is required when
the context extension adds more than one lock.

Analogously to \cref{thm:ikc-sound-possible-world}, we state the
soundness of \ISFourC with respect to \emph{reflexive and transitive}
possible\hyp{}world models before we instantiate it with the \NbE
model that we will construct in the next
\lcnamecref{sec:is4c-nbe-model}. In addition to the coherence
conditions stated before \cref{thm:ikc-sound-possible-world} the
soundness proof for \ISFourC relies on coherence conditions involving
the additional proofs~$\reflRm$ and~$\transRm$ that a reflexive and
transitive modal accessibility relation~$\Rm$ must come equipped
with. Specifically, $\transRm$ also needs to be associative, $\reflRm$
also needs to be a unit for $\transRm$, and the proofs of the
factorization condition also need to satisfy the functoriality laws in
the modal accessibility argument, \ie $\factorRi{(\reflRm{w}),i} = i$,
$\factorRm{(\reflRm{w}),i} = \reflRm{w'}$,
$\factorRi{(\transRm{n,m}),i} = \factorRi{n,i'}$ and
$\factorRm{(\transRm{n,m}),i} =
\transRm{(\factorRm{n,i'}),(\factorRm{m,i})}$ where
$i' \coloneqq \factorRi{m,i} : w \Ri w'$.

\begin{proposition}\label{pro:is4c-sound-right-adjoint-comonad}
  Let $\Cat$ be a Cartesian closed category equipped with a
  comonad~$\BoxTy$ that has a left adjoint~$\lockCtx \dashv \BoxTy$,
  then equivalent terms~$t$ and~$u : \Tm{\Gamma}{A}$ denote equal
  morphisms in $\Cat$.

  \begin{proof}
    This is a version of \textcite[Theorem~4.8]{Clouston18} for \ISFourC where
    the side condition of \cref{rule:unboxTm/ISFourC-0} appears as an
    argument to the term former~$\unboxTm$ and hence idempotency is
    not imposed on the comonad~$\BoxTy$.
  \end{proof}
\end{proposition}

\begin{theorem}\label{thm:is4c-sound-possible-world}
  Let $\Mod$ be a possible\hyp{}world
  model (see \cref{def:possible-world-model}) such that the modal
  accessibility relation~$\Rm$ is reflexive and transitive. If two
  terms~$t$ and~$u : \Tm{\Gamma}{A}$ of \ISFourC are
  equivalent (see \cref{fig:is4c-conversion})
  then the functions~$\eval{t}$ and~$\eval{u} : \forall w.\, \eval{\Gamma}_{w} \to \eval{A}_{w}$ as determined by $\Mod$ are
  equal.

  \begin{proof}
    The right adjoint determined by a reflexive and transitive frame
    has a comonad structure so that we can conclude by applying
    \cref{pro:is4c-sound-right-adjoint-comonad}.
  \end{proof}
\end{theorem}

\subsubsection{\texorpdfstring{\NbE}{NbE} Model}\label{sec:is4c-nbe-model}
The normal forms of \ISFourC are defined as before, except for the following rule
replacing the neutral rule~\labelcref{rule:unboxNe/IKC}.
\begin{mathpar}
  \inferrule[\ISFourC/Ne/$\BoxTy$\nbhyp{}Elim]{%
    \Delta \vdashNe n : \BoxTy{A}\\
    e : \RISFourC{\Delta}{\Gamma}
  }{%
    \Gamma \vdashNe \unboxISFourTm{n,e} : A
  }\label{rule:unboxNe/ISFourC}%
\end{mathpar}

\todowarn[inline]{}
The \NbE model construction also proceeds in the same way, where we now pick the
relation~$\Rm$ as arbitrary extensions of a context:
$\Delta \Rm \Gamma = \RISFourC{\Delta}{\Gamma}$.
The modal fragment for $\reify$ and $\reflect$ are now implemented as follows:
\begin{equation*}
  \begin{array}{l @{\$} l @{\;}c@{\;} l}
    \reify_{\BoxTy{A},\Gamma}   & b & = & \boxTm{(\reify_{A,(\ExtCtx{\Gamma}{\lockCtx})}{(b\$\idWk \$(\lockExt{\nilExt}))})} \\
    \reflect_{\BoxTy{A},\Gamma} & n & = & \lambda (o : \Wk{\Gamma}{\Gamma'}).\, \lambda (e : \RISFourC{\Gamma'}{\Delta}).\, \reflect_{A,\Delta}{(\unboxTm{(\wk{o,n}),e})}
  \end{array}
\end{equation*}

\begin{theorem}\label{thm:is4c-normalization}
  Let $\Mod$ denote the possible\hyp{}world model over the reflexive
  and transitive frame given by the relations~$\Wk{\Gamma}{\Gamma'}$
  and~$\RISFourC{\Delta}{\Gamma}$ and the
  valuation~$V_{\BaseTy,\Gamma} = \Ne{\Gamma}{\BaseTy}$.

  There is a
  function~$\quote : \Mod(\eval{\Gamma},\eval{A}) \to \Nf{\Gamma}{A}$
  such that the
  composite~$\norm = \quote \circ \eval : \Tm{\Gamma}{A} \to
  \Nf{\Gamma}{A}$ from terms to normal forms of \ISFourC is complete
  and adequate.
\end{theorem}
The proof of this \lcnamecref{thm:is4c-normalization} requires us to
identify terms by extending the equational theory of \ISFourC with an
additional rule.
To understand the need for it, consider unboxing a
term~$\Gamma \vdash t : \BoxTy{A}$ into an extended
context~$\ExtCtx{\Gamma}{B}$ in \ISFourC.
We may first weaken $t$ as
$\ExtCtx{\Gamma}{B} \vdash \wk{(\dropWk{\idWk}),t} : \BoxTy{A}$ and
then apply $\unboxTm$ as
$\ExtCtx{\Gamma}{B} \vdash \unboxTm{(\wk{(\dropWk{\idWk}),t}),\nilExt}
: A$.
However, we may also apply $\unboxTm$ on $t$ as
$\ExtCtx{\Gamma}{B} \vdash \unboxTm{t,(\varExt{\nilExt})} : A$.
This weakens the term ``explicitly'' in the sense that the weakening
with $B$ is recorded in the term by the proof~$\varExt{\nilExt}$ of
the modal accessibility
relation~$\RISFourC{\Gamma}{\ExtCtx{\Gamma}{B}}$.
\todowarn{}{%
  The two ways of unboxing $\Gamma \vdash t : \BoxTy{A}$ into the
  extended context~$\ExtCtx{\Gamma}{B}$ result in two terms with the
  same denotation in the possible\hyp{}world semantics but
  \emph{distinct} typing derivations.
  We wish the two typing derivations~$\unboxTm{t,(\varExt{\nilExt})}$
  and~$\unboxTm{(\wk{(\dropWk{\idWk}),t}),\nilExt}$ to be identified.
}%
For this reason, we extend the equational theory of \ISFourC with the
rule~$\unboxTm{t,(\transRm{e,e'})} \thyeq \unboxTm{(\wk{(\toOPE
    e),t}),e'}$ for any \emph{lock\hyp{}free} extension~$e$, which can
be converted to a sequence of $\dropWk$s using the function~$\toOPE$.
Explicit weakening can also be avoided by, instead of extending the
equational theory, changing the definition of the modal accessibility
relation such that $\RISFourC{\Delta}{\Gamma}$ holds only if
$\Gamma = \Delta$ or
$\Gamma = \ExtCtx{\ExtCtx{\Delta}{\lockCtx}}{\Gamma'}$ for some
$\Gamma'$.
Note that the modal accessibility relation for \IKC, where the issue
of explicit weakening does not occur, satisfies this property.

\subsection{Extending to the Calculi \texorpdfstring{\ITC}{ITC} and \texorpdfstring{\IKFourC}{IK4C}}

The \NbE model construction for \ITC and \IKFourC follows a similar
pursuit as \ISFourC.
We define suitable modal accessibility relations $\RITC{}{}$ and $\RIKFourC{}{}$
as extensions that allow the addition of at most one $\lockCtx$,
and at least one lock $\lockCtx$, respectively.
To give possible\hyp{}world semantics, we require an additional frame condition
that the relation $\Rm$ be reflexive for \ITC and transitive
for \IKFourC.
For evaluation, we use a function~$\trimITSubS : \eval{\Gamma}_w \to  \RITC{\Delta}{\Gamma} \to \eval{\ExtCtx{\Delta}{\lockCtx}}_w$ for \ITC, and similarly $\trimIKFourSubS$ for \IKFourC.
The modification to the neutral rule~\labelcref{rule:unboxNe/IKC} is achieved as before
in \ISFourC using the corresponding modal accessibility relations.
Unsurprisingly, reification and reflection can also be implemented,
thus yielding normalization functions for both \ITC and \IKFourC.

\section{Completeness, Decidability and Logical Applications}\label{sec:corollaries}

In this \lcnamecref{sec:corollaries} we record some immediate
consequences of the model constructions we presented in the previous
\lcnamecref{sec:nbe-models}.

\def\MM{\Mod_{0}}
\def\NN{\Mod[N]}

\paragraph{Completeness of the Equational Theory}
As a corollary of the adequacy of an \NbE model~$\NN$, \ie
$\Gamma \vdash t \thyeq u : A$ whenever
$\eval{t} = \eval{u} : \NN(\eval{\Gamma},\eval{A})$, we obtain
completeness of the equational theory with respect to the class of
models that the respective \NbE model belongs to. Given the \NbE
models constructed in \cref{sec:ikc-nbe-model,sec:is4c-nbe-model} this
means that the equational theories of \IKC and
\ISFourC (\cf \cref{fig:ikc-conversion-full}) are (sound and) complete
with respect to the class of Cartesian closed categories equipped with
an adjunction and a right\hyp{}adjoint comonad, respectively.

\begin{theorem}\label{thm:ikc-categorically-complete}
  Let $t$, $u : \Tm{\Gamma}{A}$ be two terms of \IKC. If for all
  Cartesian closed categories~$\Mod$ equipped with an adjunction it is
  the case that $\eval{t} = \eval{u} : \Mod(\eval{\Gamma},\eval{A})$
  then $\Gamma \vdash t \thyeq u : A$.

  \begin{proof}
    Let~$\MM$ be the model we constructed in
    \cref{sec:ikc-nbe-model}. Since $\MM$ is a Cartesian closed
    category equipped with an adjunction, by assumption we have
    $\eval{t}_{\MM} = \eval{u}_{\MM}$. And lastly, since $\MM$ is an
    \NbE model, we have
    $\Gamma \vdash t \thyeq \quote{\eval{t}_{\MM}} =
    \quote{\eval{u}_{\MM}} \thyeq u : A$.
  \end{proof}
\end{theorem}

Note that this statement corresponds to
\textcite[Theorem~3.2]{Clouston18} but there it is obtained via a term
model construction and for the term model \todohint{}{to be equipped with an
adjunction} the calculus needs to be first extended with an
internalization of the operation~$\lockCtx$ on contexts as an
operation~$\lockTy$ on types.
\todowarn[inline]{}

\begin{theorem}\label{thm:is4c-categorically-complete}
  Let $t$, $u : \Tm{\Gamma}{A}$ be two terms of \ISFourC. If for all
  Cartesian closed categories~$\Mod$ equipped with a
  right\hyp{}adjoint comonad
  it is the case that
  $\eval{t} = \eval{u} : \Mod(\eval{\Gamma},\eval{A})$ then
  $\Gamma \vdash t \thyeq u : A$.

  \begin{proof}
    As for \cref{thm:ikc-categorically-complete}.
  \end{proof}
\end{theorem}

This statement corresponds to \textcite[Section~4.4]{Clouston18} but
there it is proved for an equational theory that identifies terms up
to differences in the accessibility proofs and with respect to the
class of models where the comonad is \emph{idempotent}, to which the
model of \cref{sec:is4c-nbe-model} does not belong.

\paragraph{Completeness of the Deductive Theory}
Using the quotation function of an \NbE model~$\NN$, \ie
$\quote : \NN(\eval{\Gamma},\eval{A}) \to \Tm{\Gamma}{A}$, we obtain
completeness of the deductive theory with respect to the class of
models that the respective \NbE model belongs to. Given the \NbE
models constructed in \cref{sec:ikc-nbe-model,sec:is4c-nbe-model} this
means that the deductive theories of \IKC and
\ISFourC (\cf \cref{fig:ikc-syn-full,fig:it-ik4-is4-syn}) are (sound
and) complete with respect to the class of possible\hyp{}world models
with an arbitrary frame and a reflexive\ndash{}transitive frame,
respectively.

\begin{theorem}\label{thm:ikc-logically-complete}
  Let $\Gamma : \Ctx$ be a context and $A : \Ty$ a
  type. If for all possible\hyp{}world models $\Mod$ it is the case
  that \todowarn{}{$\Mod(\eval{\Gamma},\eval{A})$ is inhabited} then
  there is a
  term~$t : \Tm{\Gamma}{A}$ of \IKC.

  \begin{proof}
    Let $\MM$ be the model we constructed in
    \cref{sec:ikc-nbe-model}. Since $\MM$ is a possible\hyp{}world
    model, by assumption we have a
    morphism~$p : \MM(\eval{\Gamma},\eval{A})$. And lastly, since
    $\MM$ is an \NbE model, we have the
    term~$\quote{p} : \Tm{\Gamma}{A}$.
  \end{proof}
\end{theorem}

\begin{theorem}\label{thm:is4c-logically-complete}
  Let $\Gamma : \Ctx$ be a context and $A : \Ty$ a
  type. If for all possible\hyp{}world models $\Mod$ with a
  reflexive\ndash{}transitive frame it is the case that
  $\Mod(\eval{\Gamma},\eval{A})$ is inhabited then there is a
  term~$t : \Tm{\Gamma}{A}$ of \ISFourC.

  \begin{proof}
    As for \cref{thm:ikc-logically-complete}.
  \end{proof}
\end{theorem}

Note that the proofs of
\cref{thm:ikc-logically-complete,thm:is4c-logically-complete} are
constructive.
\todowarn[inline]{}

\todohint[inline]{}

\paragraph{Decidability of the Equational Theory}
As a corollary of the completeness and adequacy of an \NbE
model~$\NN$, \ie $\Gamma \vdash t \thyeq u : A$ if and only if
$\eval{t} = \eval{u} : \NN(\eval{\Gamma},\eval{A})$, we obtain
decidability of the equational theory from decidability of the
equality of normal forms~$n$,~$m : \Nf{\Gamma}{A}$. Given the \NbE
models constructed in \cref{sec:ikc-nbe-model,sec:is4c-nbe-model}
this means
that the equational theories of \IKC and
\ISFourC (\cf \cref{fig:ikc-conversion-full}) are decidable.

To show that any of the following decision problems~$P(x)$ is
decidable we give a \emph{constructive} proof of the
proposition~$\forall x.\, P(x) \lor \neg P(x)$. Such a proof can be
understood as the construction of an algorithm~$d$ that takes as input
an~$x$ and produces as output a Boolean~$d(x)$, alongside a
correctness proof that $d(x)$ is $\trueTm$ if and only if $P(x)$
holds.

\begin{theorem}\label{thm:ikc-conversion-decidable}
  For any two terms~$t$,~$u : \Tm{\Gamma}{A}$ of \IKC the problem
  whether $t \thyeq u$ is decidable.

  \begin{proof}
    We first observe that
    for any two normal forms~$n$,~$m : \Nf{\Gamma}{A}$ of \IKC the
    problem whether $n = m$ is decidable by proving
    $\forall n, m.\, n = m \lor n \neq m$ constructively. All the
    cases of an simultaneous induction on $n$, $m : \Nf{\Gamma}{A}$
    are immediate.

    Let $\NN$ be the \NbE model we constructed in
    \cref{sec:ikc-nbe-model}. Completeness and adequacy of $\NN$ imply
    that we have $t \thyeq u$ if and only if $\norm{t} = \norm{u}$ for
    the function~$\norm : \Tm{\Gamma}{A} \to \Nf{\Gamma}{A}, t \mapsto
    \quote{\eval{t}}$. Now, $t \thyeq u$ is decidable because
    $\norm{t} = \norm{u}$ is decidable by
    the observation we started with%
    .
  \end{proof}
\end{theorem}

\begin{theorem}\label{thm:is4c-conversion-decidable}
  For any two terms~$t$,~$u : \Tm{\Gamma}{A}$ of \ISFourC the problem
  whether $t \thyeq u$ is decidable.

  \begin{proof}
    As for \cref{thm:ikc-conversion-decidable}.
  \end{proof}
\end{theorem}

\paragraph{Denecessitation}
The last of the consequences of the \NbE model constructions we record
is of a less generic flavour than the other three, namely it is an
application of normal forms to a basic proof\hyp{}theoretic result in
modal logic.

Using invariance of truth in possible\hyp{}world models under
bisimulation%
\footnote{\todowarn[inline]{}{Invariance of
    truth under bisimulation} says that if $w$ and $v$ are two
  bisimilar worlds in two possible\hyp{}world models $\Mod_{0}$ and
  $\Mod_{1}$, respectively, then for all formulas~$A$ it is the case
  that $\eval{A}_{w}$ holds in $\Mod_{0}$ if and only if
  $\eval{A}_{v}$ does in $\Mod_{1}$.}
it can be shown that $\BoxTy{A}$ is a valid formula of \IK (or
\ISFour) if and only if $A$ is. A completeness theorem then implies
the same for provability of $\BoxTy{A}$ and $A$. The statement for
proofs in \IKC (and \ISFourC) can also be shown by inspection of
normal forms as follows.

Firstly, we note that while deduction is not closed under arbitrary
context extensions (including locks) it is closed under extensions
(including locks) \emph{on the left}:
\begin{lemma}[{\cf \textcite[Lemma~A.1]{Clouston18}}]\label{lem:left-weakening}
  Let $\Delta$, $\Gamma : \Ctx$ be arbitrary contexts, both possibly containing locks, and $A : \Ty$ an
  arbitrary type. There is an
  operation~$\Tm{\Gamma}{A} \to \Tm{\ProdCtx{\Delta}{\Gamma}}{A}$ on
  terms of \IKC (and \ISFourC), where $\ProdCtx{\Delta}{\Gamma}$
  denotes context concatenation.

  \begin{proof}
    By recursion on terms.
  \end{proof}
\end{lemma}

And, secondly, we note that also a converse of this
\lcnamecref{lem:left-weakening} holds by inspection of normal forms:
\begin{lemma}\label{lem:left-unweakening}
  Let $\Delta$, $\Gamma : \Ctx$ be arbitrary contexts, both possibly containing locks, $A : \Ty$ an
  arbitrary type and $t : \Tm{\ProdCtx{\Delta}{\Gamma}}{A}$ a term of
  \IKC (or \ISFourC) in the concatenated
  context~$\ProdCtx{\Delta}{\Gamma}$ that does \emph{not} mention any
  variables from $\Delta$, then there is a term~$t' : \Tm{\Gamma}{A}$
  of \IKC (or \ISFourC, respectively)%
  .

  \begin{proof}
    Since
    normalization (see \cref{thm:ikc-normalization,thm:is4c-normalization})
    does not introduce new free variables it suffices to prove the
    statement for terms in normal form. We do so by induction on
    normal
    forms~$n :
    \Nf{\ProdCtx{\Delta}{\Gamma}}{A}$ (see \cref{fig:ikc-nf}). The
    only nonimmediate step is for $n$ of the form~$\unboxTm{n',e}$ for
    some neutral element~$n' : \Ne{\Delta'}{\BoxTy{A}}$ and
    $\R{\Delta'}{\Wk{\Delta}{\ProdCtx{\Delta}{\Gamma}}}$. But in that
    case the induction hypothesis says that we have a neutral
    element~$n'' : \Ne{\EmptyCtx}{\BoxTy{A}}$, which is impossible.
  \end{proof}
\end{lemma}

Note that some form of normalization seems to be needed in the proof
of \cref{lem:left-unweakening}. More specifically, the
``strengthening'' of a term of the form~$\unboxTm{t,e}$ from the
context~$\ProdCtx{\ExtCtx{\EmptyCtx}{\lockCtx}}{\EmptyCtx}$ to the
empty context~$\EmptyCtx$ cannot possibly result in a term of the
form~$\unboxTm{t',e'}$ because there is \emph{no} context~$\Gamma$
such that $\R{\Gamma}{\EmptyCtx}$ in \IKC. As an example, consider the
term~$\unboxTm{(\boxTm{(\lamTm{x.\,x})}),\nilExt}$, which needs to be
strengthened to $\lamTm{x.\,x}$.

With these two \lcnamecrefs{lem:left-weakening} at hand we are ready
to prove denecessitation through normalization:
\begin{theorem}\label{thm:denecessitation}
  Let $A : \Ty$ be an arbitrary type. There is a
  term $t : \Tm{\EmptyCtx}{A}$ of \IKC (or \ISFourC) if and only if there
  is a
  term $u : \Tm{\EmptyCtx}{\BoxTy{A}}$ of \IKC (or \ISFourC,
  respectively), where $\EmptyCtx : \Ctx$ denotes the \emph{empty}
  context.

  \begin{proof}
    From a term~$t : \Tm{\EmptyCtx}{A}$ we can construct a
    term~$t' : \Tm{\ExtCtx{\EmptyCtx}{\lockCtx}}{A}$ using
    \cref{lem:left-weakening} and thus the
    term~$u = \boxTm{t'} : \EmptyCtx \vdash \BoxTy{A}$.

    In the other direction, from a
    term~$u : \Tm{\EmptyCtx}{\BoxTy{A}}$ we obtain a normal
    form~$u' = \norm{u} : \Nf{\EmptyCtx}{\BoxTy{A}}$ using
    \cref{thm:ikc-normalization,thm:is4c-normalization}. By inspection
    of normal forms (see \cref{fig:ikc-nf}) we know that $u'$ must be
    of the form~$\boxTm{v}$ for some normal
    form~$v : \Nf{\ExtCtx{\EmptyCtx}{\lockCtx}}{A}$, from which we
    obtain a term~$t : \Tm{\EmptyCtx}{A}$ using
    \cref{lem:left-unweakening} since the
    context~$\ExtCtx{\EmptyCtx}{\lockCtx}$ does not declare any
    variables that could have been mentioned in $v$.
  \end{proof}
\end{theorem}

This concludes this \lcnamecref{sec:corollaries} on some consequences
of the model constructions presented in this paper. Note that the
consequences we recorded are completely independent of the concrete
model construction. To wit, the two completeness theorems follow from
the mere existence of an \NbE model, and the decidability and
denecessitation theorems follow from the mere existence of a
normalization function.

\section{Programming\texorpdfstring{\hyp{}}{-}Language Applications}\label{sec:applications}

In this \lcnamecref{sec:applications}, we discuss some implications
of normalization for Fitch\hyp{}style calculi for specific
interpretations of the necessity modality in the context of
programming languages.
In particular, we show how normalization can be used to prove properties
about program calculi by leveraging the shape of normal forms of terms.
We extend the term calculi presented earlier with
application\hyp{}specific primitives, ensure that the extended calculi
are in fact normalizing, and then use this result to prove properties
such as capability safety, noninterference, and binding\hyp{}time
correctness.
Note that we do not mechanize these results in \Agda{} and do not
prove these properties in their full generality, but only illustrate
special cases.
Although possible, proving the general properties requires further
technical development that obscures the main idea underlying the use
of normal forms for simplifying these proofs.

\subsection{Capability Safety}\label{sec:capability-safety}

\Textcite{ChoudhuryK20} present a modal type system based on \ISFour
for a programming language with implicit effects in the style of
\ML~\parencite{MilnerTH90} and the computational lambda
calculus~\parencite{Moggi89b}.
In this language, programs need access to capabilities to perform
effects.
For instance, a primitive for printing a string requires a capability
as an argument in addition to the string to be printed.
Crucially, capabilities cannot be introduced within the language, and
must be obtained either from the global context (called \emph{ambient}
capabilities) or as a function argument.

Let us denote the type of capabilities by $\CapTy$.
Passing a printing capability~$c$ to a function of
type~$\CapTy \FunTy \UnitTy$ in a language that uses capabilities to
print yields a program that either
\begin{enumerate*}
\item does not print,\label{en:one}
\item prints using only the capability~$c$, or\label{en:two}
\item prints using ambient capabilities (and possibly
  $c$).\label{en:three}
\end{enumerate*}
A program that at most uses the capabilities that it is passed
explicitly, as in the cases~\labelcref{en:one} and~\labelcref{en:two}, is said to be
\emph{capability safe}.
To identify such programs, \textcite{ChoudhuryK20} introduce a
comonadic modality~$\BoxTy$ to capture capability safety.
Their type system is loosely based on the
dual\hyp{}context calculus for \ISFour~\parencite{PfenningD01,Kavvos20}.
A term of type~$\BoxTy{A}$ is enforced to be capability safe by making
the introduction rule for $\BoxTy$ ``brutally'' remove all
capabilities from the typing context.
As a result, programs with the type~$\BoxTy{(\CapTy \FunTy \UnitTy)}$
are denied ambient capabilities and thus guaranteed to behave like
the cases~\labelcref{en:one} and~\labelcref{en:two}.

\Textcite{ChoudhuryK20} characterize capability safety precisely using
their \emph{capability space} model.
A capability space~$\pair{X}{w_X}$ is a set~$X$ and a weight relation~$w_X$
that assigns sets of capabilities to every member in $X$.
In this model, they define a comonad that restricts the underlying set
of a capability space to those elements that are only related to the
empty set of capabilities.
This comonad has a left adjoint that replaces the weight relation of a
capability space by the relation that relates every element to the
empty set of capabilities.
This adjunction suggests that capability spaces are a model of \ISFourC and
we may thus use \ISFourC to write programs that support reasoning
about capability safety.

In this \lcnamecref{sec:capability-safety}, we present a calculus~\ISFourCMoggiCap{}
that extends \ISFourC with a capability type and a
monad for printing effects.
We extend the normalization algorithm for \ISFourC to
\ISFourCMoggiCap{} and show that the resulting normal forms can be
used to prove a kind of capability safety.
In contrast to the language presented by \textcite{ChoudhuryK20},
\ISFourCMoggiCap{} models a language where effects are explicit in the
type of a term.
Languages with explicit effects, such as \Haskell~\parencite{Haskell}
(with the $\IOTy$ monad) or \PureScript~\parencite{PureScript} (with
the \mintinline{Haskell}{Effect} monad), can also benefit from a
mechanism for capability safety, and we begin with an example in a
hypothetical extension of \PureScript to illustrate this.

\paragraph{Example in \PureScript} Let us consider web development in
\PureScript.
A web application may consist of a mashup of several components, \eg
social media, news feed, or chat, provided by untrusted sources.
A component is a function of type
\begin{minted}{Haskell}
  type Component = Element -> Effect Unit
\end{minted}
that takes as a parameter the \DOM element where the component will be
rendered.
For the correct functioning of the web application, it is important that
components do not interfere with each other in malicious ways.
For example, a malicious component (of Bob) could illegitimately
overwrite a \DOM element (of Alice):
\begin{minted}{Haskell}
  evilBob :: Component
  evilBob e = do w      <- window
                 doc    <- document w
                 aliceE <- getElementById "alice.app" doc
                 setTextContent "Alice has been hacked!" aliceE
\end{minted}

The issue here is that Bob has unrestricted access to the function~\mintinline{Haskell}{window :: Effect Window},
and is able to obtain
the \DOM using \mintinline{Haskell}{document :: Window -> Effect DOM}
and overwrite an element that belongs to Alice.
Capabilities can be leveraged to restrict the access to
\mintinline{Haskell}{window}.
We can achieve this by extending \PureScript with a
type~\mintinline{Haskell}{WindowCap}, a type
constructor~\mintinline{Haskell}{Box} that works similarly to
\citeauthor{ChoudhuryK20}'s $\BoxTy$, and replacing the
function~\mintinline{Haskell}{window} with a
function~\mintinline{Haskell}{window' :: WindowCap -> Effect Window}
that requires an additional capability argument.
By making \mintinline{Haskell}{WindowCap} an ambient capability that
is available globally, all existing programs retain their unrestricted
access to retrieve a window as before.
The difference now, however, is that we can selectively restrict some
programs and limit their access to \mintinline{Haskell}{WindowCap}
using \mintinline{Haskell}{Box}.
We can define a variant of the type~\mintinline{Haskell}{Component}
as:
\begin{minted}{Haskell}
  type Component' = Box (Element -> Effect Unit)
\end{minted}

By requiring Bob to write a component of the type~\mintinline{Haskell}{Component'},
we are ensured that Bob cannot
overwrite an element that belongs to Alice.
This is because the \mintinline{Haskell}{Box} type constructor used to
define \mintinline{Haskell}{Component'} disallows access to all
ambient capabilities (including \mintinline{Haskell}{WindowCap}), and
thus restricts Bob to only using the given
\mintinline{Haskell}{Element} argument.
In particular, the program~\mintinline{Haskell}{evilBob} cannot be
reproduced with the type~\mintinline{Haskell}{Component'} since the
substitute function~\mintinline{Haskell}{window'} requires a
capability that is neither available as an argument nor as an ambient
capability.
\begin{figure}
  \begin{align*}
    \Ty\quad A, B \Coloneqq \ldots\ |\ \TTy{A}\ |\ \CapTy\ |\ \StringTy\ |\ \UnitTy & &
    \Ctx\quad \Gamma \Coloneqq \ldots
  \end{align*}
  \begin{mathpar}
    \inferrule[$\TTy$\nbhyp{}Intro]{%
      \Gamma  \vdash t : A
    }{%
      \Gamma \vdash \returnTm{t} : \TTy{A}
    }

    \inferrule[$\TTy$\nbhyp{}Elim]{%
      \Gamma  \vdash t : \TTy{A}\\
      \ExtCtx{\Gamma}{A} \vdash u : \TTy{B}
    }{%
      \Gamma \vdash \letTm{t,u} : \TTy{B}
    }

    \\

    \inferrule[$\UnitTy$\nbhyp{}Intro]{%
    }{%
      \Gamma \vdash \unitTm : \UnitTy
    }

    \inferrule[$\StringTy$\nbhyp{}Lit]{ }
    [s \in \StringSet]{%
      \Gamma \vdash \strTm{s} : \StringTy
    }

    \\

    \inferrule[$\TTy$\nbhyp{}Print]{%
      \Gamma  \vdash c : \CapTy\\
      \Gamma  \vdash s : \StringTy
    }{%
      \Gamma \vdash \printTm{c,s} : \TTy{\UnitTy}
    }
  \end{mathpar}
  \caption{Types, contexts and terms of \ISFourCMoggiCap (omitting the
    unchanged rules of \cref{fig:ikc-syn-full,fig:is4-full})}
  \label{fig:is4c-moggi-full}
\end{figure}

\begin{figure}[ht]
  \begin{mathpar}
    \inferrule[$\TTy$\nbhyp{}$\beta$]{ %
      \Gamma \vdash t : A\\
      \ExtCtx{\Gamma}{A} \vdash u : \TTy{B}
    }{%
      \Gamma \vdash \letTm{(\returnTm{t}),u} \thyeq \subst{(\extTmSub{\idSub,t}),u}
    }%

    \inferrule[$\TTy$\nbhyp{}$\eta$]{ %
      \Gamma \vdash t : \TTy{A}
    }{%
      \Gamma \vdash t \thyeq \letTm{t,(\returnTm{(\varTm{\zeroVar}}))}
    }%

    \inferrule[$\TTy$\nbhyp{}$\gamma$]{ %
      \Gamma \vdash t_1 : A\\
      \ExtCtx{\Gamma}{A} \vdash t_2 : \TTy{B}\\
      \ExtCtx{\Gamma}{B} \vdash t_3 : \TTy{C}
    }{%
    \Gamma \vdash \letTm{(\letTm{t_1,t_2}),t_3} \thyeq \letTm{t_1,(\letTm{t_2,(\wk{(\keepVarWk{(\dropIdWk)}),t_3}}))}
    }%
  \end{mathpar}
  \caption{Equational theory for \ISFourCMoggiCap (omitting the unchanged
    rules of \cref{fig:ikc-conversion-full,fig:is4c-conversion})}
  \label{fig:is4c-moggi-conversion-full}
\end{figure}

\begin{figure}[ht]
  \begin{mathpar}
    \inferrule[Nf/$\TTy$\nbhyp{}Intro]{%
      \Gamma  \vdashNf m : A
    }{%
      \Gamma \vdashNf \returnTm{m} : \TTy{A}
    }

    \inferrule[Nf/$\TTy$\nbhyp{}Elim]{%
      \Gamma  \vdashNe n : \TTy{A}\\
      \ExtCtx{\Gamma}{A} \vdashNf m : \TTy{B}
    }{%
      \Gamma \vdashNf \letTm{n,m} : \TTy{B}
    }

    \\

    \inferrule[Nf/$\UnitTy$\nbhyp{}Intro]{%
    }{%
      \Gamma \vdashNf \unitTm : \UnitTy
    }

    \inferrule[Nf/Up\nbhyp{}$\CapTy$]{
      \Gamma \vdashNe n : \CapTy
    }
    {%
      \Gamma \vdashNf \upTm{n} : \CapTy
    }

    \inferrule[Nf/Up\nbhyp{}$\StringTy$]{
      \Gamma \vdashNe n : \StringTy
    }
    {%
      \Gamma \vdashNf \upTm{n} : \StringTy
    }

    \inferrule[Nf/$\StringTy$\nbhyp{}Lit]{ }
    [s \in \StringSet]{%
      \Gamma \vdashNf \strTm{s} : \StringTy
    }

    \\

    \mprset{sep=1em}
    \inferrule[Nf/$\TTy$\nbhyp{}Print]{%
      \Gamma  \vdashNf c : \CapTy\\
      \Gamma  \vdashNf s : \StringTy\\
      \ExtCtx{\Gamma}{\UnitTy} \vdashNf m : \TTy{A}
    }{%
      \Gamma \vdashNf \letTm{(\printTm{c,s}),m} : \TTy{A}
    }
  \end{mathpar}
  \caption{Normal forms of \ISFourCMoggiCap (omitting the unchanged
    normal forms of \ISFourC)}
  \label{fig:is4c-moggi-nf}
\end{figure}

\paragraph{Extension with a Capability and a Monad}
We extend \ISFourC with a monad for printing based on Moggi's monadic
metalanguage~\parencite{Moggi91}.
We introduce a type~$\TTy{A}$ that denotes a monadic computation that
can print before returning a value of type~$A$, a type~$\CapTy$ for
capabilities, and a type~$\StringTy$ for strings.
\Cref{fig:is4c-moggi-full} summarizes the terms that correspond to
this extension.
The term construct~$\printTm$ is used for printing.
The equational theory of \ISFourCMoggiCap and the corresponding normal
forms are summarized in \cref{fig:is4c-moggi-conversion-full} and
\cref{fig:is4c-moggi-nf}, respectively.

To extend the \NbE model of \ISFourC with an interpretation for the
monad, we use the standard techniques used for normalizing
computational
effects~\parencite{AhmanS13,Filinski01}.
The interpretation of the other primitive types also follows a
standard pursuit~\parencite{ValliappanRL21}: we interpret $\CapTy$ by
neutrals of type~$\CapTy$ and $\StringTy$ by the disjoint union of
$\StringSet$ and neutrals of type~$\StringTy$.
The difference in their interpretation is caused by the fact that
there is no introduction form for the type~$\CapTy$.

\paragraph{Proving Capability Safety}
Programs that lack access to capabilities are necessarily capability
safe.
We say that a program~$\Gamma \vdash p : A$ is \emph{trivially
  capability safe} if there is a program~$\EmptyCtx \vdash p' : A$
such that $\Gamma \vdash p \thyeq \leftConcat_\Gamma{p'} : A$, where
$\leftConcat_\Gamma : \forall \Delta, A.\, \Tm{\Delta}{A} \to
\Tm{\ProdCtx{\Gamma}{\Delta}}{A}$
can be defined similarly to the operation given by
\cref{lem:left-weakening} for \ISFourC.

First, we prove an auxiliary \lcnamecref{lem:nf-caplock} about normal
forms with a capability in context.
\begin{lemma}\label{lem:nf-caplock}
  For any context~$\Gamma$, type~$A$ and normal
  form~$\ProdCtx{\ExtCtx{c : \CapTy}{\lockCtx}}{\Gamma} \vdashNf n :
  A$ there is a normal
  form~$\ProdCtx{\ExtCtx{\EmptyCtx}{\lockCtx}}{\Gamma} \vdashNf n' :
  A$ such that $n = \leftConcat_{c : \CapTy}{n'}$.

  \begin{proof}
    We prove the statement for both normal forms and neutral elements
    by mutual induction.
    The only nonimmediate case is when the neutral is of the
    form~$\ProdCtx{\ExtCtx{c : \CapTy}{\lockCtx}}{\Gamma} \vdashNe
    \unboxTm{n,e} : A$ for some $n : \Delta \vdashNe \BoxTy{A}$ and
    $e : \RISFourC{\Delta}{\ProdCtx{\ExtCtx{c :
          \CapTy}{\lockCtx}}{\Gamma}}$.
    We observe that there are no neutral elements of type~$\BoxTy{A}$
    in context~$c : \CapTy$ and that hence $\Delta$ must contain the
    leftmost lock in
    $\ProdCtx{\ExtCtx{c : \CapTy}{\lockCtx}}{\Gamma}$.
    Thus, this case also holds by induction hypothesis.
  \end{proof}
\end{lemma}

Now, we observe that all terms~$c : \CapTy \vdash t : \BoxTy{A}$ are
trivially capability safe.
By normalization, we have that
$c : \CapTy \vdash t \thyeq \norm{t} : \BoxTy{A}$.
Given the definition of normal forms of \ISFourCMoggiCap, $\norm{t}$
must be $\boxTm{n}$ for some normal
form~$\ExtCtx{c : \CapTy}{\lockCtx} \vdashNf n : A$.
By \cref{lem:nf-caplock}, there is a normal
form~$\lockCtx \vdashNf n' : A$ such that
$n = \leftConcat_{\ExtCtx{\EmptyCtx}{\CapTy}}{n'}$.
Since the operation $\leftConcat$ commutes with $\boxTm$, \ie
$\leftConcat_{\ExtCtx{\EmptyCtx}{\CapTy}}{(\boxTm{n'})} =
\boxTm{(\leftConcat_{\ExtCtx{\EmptyCtx}{\CapTy}}{n'})}$, we also have
that
$t \thyeq \boxTm{n} =
\leftConcat_{\ExtCtx{\EmptyCtx}{\CapTy}}{(\boxTm{n'})}$.
As a result, $t$ must be trivially capability safe.

A consequence of this observation is that any
term~$c : \CapTy \vdash t : \BoxTy{(\TTy{\UnitTy})}$ is trivially
capability safe.
This means that $t$ does not print since it could not possibly do so
without a capability.
Going further, we can also observe that
$t \thyeq \boxTm{(\returnTm{\unitTm})} : \BoxTy{(\TTy{\UnitTy})}$, since the only normal
form of type~$\TTy{\UnitTy}$ in the empty context is
$\EmptyCtx \vdashNf \returnTm{\unitTm} : \TTy{\UnitTy}$.
Note that this argument (and the one above) readily adapts to a vector
of capabilities~$\vec{c}$ in context as opposed to a single
capability~$c$.

\subsection{Information\texorpdfstring{\hyp{}}{-}Flow Control}\label{sec:ifc}

Information\hyp{}flow control~(\IFC)~\parencite{SabelfeldM03} is a
technique used to protect the confidentiality of data in a program by
tracking the flow of information within the program.

In type\hyp{}based \emph{static}
\IFC{}~\parencite[\eg{}][]{AbadiBHR99,ShikumaI08,RussoCH08} types are
used to associate values with confidentiality levels such as
\emph{secret} or \emph{public}.
The type system ensures that secret inputs do not interfere with
public outputs, enforcing a security policy that is typically
formalized as a kind of \emph{noninterference}
property~\parencite{GoguenM82a}.

Noninterference is proved by reasoning about the semantic behaviour of
a program.
\Textcite{TomeV19} present a proof technique that uses normalization
for showing noninterference for a static \IFC{} calculus based on Moggi's
monadic metalanguage~\parencite{Moggi91}.
This technique exploits the insight that normal forms represent
equivalence classes of terms identified by their semantics, and thus
reasoning about normal forms of terms (as opposed to terms themselves)
vastly reduces the set of programs that we must take into
consideration.
Having developed normalization for Fitch\hyp{}style calculi, we can
leverage this technique to prove noninterference.

In this \lcnamecref{sec:ifc}, we extend \IKC with Booleans (denoted
\IKCBool), extend the \NbE model of \IKC to \IKCBool, and illustrate
the technique of \citeauthor{TomeV19} on \IKCBool for proving
noninterference.
We interpret the type~$\BoxTy{A}$ as a secret of type~$A$, and other
types as public.

\paragraph{Extension with Booleans} Noninterference can be better
appreciated in the presence of a type whose values are distinguishable
by an external observer.
To this extent, we extend \IKC with a type~$\BoolTy$ and corresponding
introduction and elimination forms\mdash{}as described in
\cref{fig:boolc-full}.

\begin{figure}[H]
  \begin{align*}
    \Ty\quad A, B \Coloneqq \ldots\ |\ \BoolTy & &
    \Ctx\quad \Gamma \Coloneqq \ldots
  \end{align*}
  \begin{mathpar}
    \inferrule[$\BoolTy$\nbhyp{}Intro\nbhyp{}$\trueTm$]{%
    }{%
      \Gamma \vdash \trueTm : \BoolTy
    }\label{rule:Bool-intro-true/BoolC}

    \inferrule[$\BoolTy$\nbhyp{}Intro\nbhyp{}$\falseTm$]{%
    }{%
      \Gamma \vdash \falseTm : \BoolTy
    }\label{rule:Bool-intro-false/BoolC}

    \inferrule[$\BoolTy$\nbhyp{}Elim]{%
      \Gamma  \vdash b : \BoolTy\\
      \ExtCtx{\Gamma}{\Gamma'}  \vdash t_1 : A\\
      \ExtCtx{\Gamma}{\Gamma'}  \vdash t_2 : A\\
    }{%
      \ExtCtx{\Gamma}{\Gamma'} \vdash \ifteTm{b,t_1,t_2} : A
    }\label{rule:Bool-elim/BoolC}%
  \end{mathpar}
  \caption{Types, contexts and intrinsically\hyp{}typed terms of
    \IKCBool (omitting the unchanged rules of \cref{fig:ikc-syn-full})}
  \label{fig:boolc-full}
\end{figure}

We modify the usual elimination rule for $\BoolTy$ by allowing the
context of the conclusion~$\ifteTm{b,t_1,t_2}$ and branches $t_1$ and
$t_2$ in the rule~\labelcref{rule:Bool-elim/BoolC} to extend the
context of the scrutinee~$b$.
This modification (following \textcite[Fig.~2]{Clouston18}) enables
the following \emph{commuting conversion}, which is required to ensure
that terms can be fully normalized and normal forms enjoy the
subformula property:
\begin{mathpar}
  \inferrule*{ %
    \Delta \vdash b : \BoolTy\\
    \ExtCtx{\Delta}{\Delta'} \vdash t_1 : \BoxTy{A}\\
    \ExtCtx{\Delta}{\Delta'} \vdash t_2 : \BoxTy{A}\\
    e : \R{\ExtCtx{\Delta}{\Delta'}}{\Gamma}
  }{%
    \Gamma \vdash \unboxTm{(\ifteTm{b,t_1,t_2}),e} \thyeq \ifteTm{b,(\unboxTm{t_1,e}),(\unboxTm{t_2,e})}
  }
\end{mathpar}
A commuting conversion is required as usual for every other
elimination rule, including the rule~\labelcref{rule:appTm}.
These are however standard and thus omitted here.

We extend the equational theory of \IKC to \IKCBool by adding the usual
rules~$\ifteTm{\trueTm,t_1,t_2} \thyeq t_1$,~$\ifteTm{\falseTm,t_1,t_2} \thyeq t_2$, and~$t \thyeq \ifteTm{t,\trueTm,\falseTm}$ for terms~$t$ of type~$\BoolTy$.
The normal forms of \IKCBool include those of \IKC in addition to the
following.
\begin{mathpar}
  \inferrule[Nf/$\BoolTy$\nbhyp{}Intro\nbhyp{}$\trueTm$]{%
  }{%
    \Gamma \vdashNf \trueTm : \BoolTy
  }\label{rule:Bool-intro-trueNf/BoolC}

  \inferrule[Nf/$\BoolTy$\nbhyp{}Intro\nbhyp{}$\falseTm$]{%
  }{%
    \Gamma \vdashNf \falseTm : \BoolTy
  }\label{rule:Bool-intro-falseNf/BoolC}

  \mprset{sep=1em}
  \inferrule[Nf/$\BoolTy$\nbhyp{}Elim]{%
    \Gamma  \vdashNe n : \BoolTy \\
    \ExtCtx{\Gamma}{\Gamma'}  \vdashNf m_1 : A \\
    \ExtCtx{\Gamma}{\Gamma'}  \vdashNf m_2 : A
  }{%
    \ExtCtx{\Gamma}{\Gamma'} \vdashNf \ifteTm{n,m_1,m_2} : A
  }\label{rule:Bool-elimNf/BoolC}%
\end{mathpar}
Observe that a neutral of type~$\BoolTy$ is not immediately in normal
form, and must be expanded as $\ifteTm{n,\trueTm,\falseTm}$.
This is unlike neutrals of the type~$\BaseTy$, which are in normal
form by \cref{rule:upNf}.

To extend the \NbE model of \IKC with Booleans, we leverage the
interpretation of sum types used by \textcite{AbelS19}, who attribute
their idea to \textcite{AltenkirchU04}.
This interpretation readily supports commuting conversions, and a
minor refinement that reflects the change to the
rule~\labelcref{rule:Bool-elim/BoolC} yields a reifiable
interpretation for Booleans in \IKCBool.

\paragraph{Proving Noninterference} A
program~$\EmptyCtx \vdash f : \BoxTy{A} \FunTy \BoolTy$ is
\emph{noninterferent} if it is the case that
$\EmptyCtx \vdash \appTm{f,s_1} \thyeq \appTm{f,s_2} : \BoolTy$
for any two secrets~$\EmptyCtx \vdash s_1, s_2 : \BoxTy{A}$.
By instantiating $A$ to $\BoolTy$, we can show that any
program~$\EmptyCtx \vdash f : \BoxTy{\BoolTy} \FunTy \BoolTy$ is
noninterferent and thus cannot leak a secret Boolean argument.
In \IKCBool, the type system ensures that data of type~$\BoxTy{A}$
type can only influence (or \emph{flow} to) data of type~$\BoxTy{B}$,
thus all programs of type~$\BoxTy{\BoolTy} \FunTy \BoolTy$ must be
noninterferent.
To show this, we analyze the possible normal forms of $f$ and observe
that they must be equivalent to a constant function, such as
$\lamTm{x.\, \trueTm}$ or $\lamTm{x.\, \falseTm}$, which evidently
does not use its input argument~$x$ and is thus noninterferent.

In detail, normal forms of type~$\BoxTy{\BoolTy} \FunTy \BoolTy$
must have the shape~$\lamTm{x.\, m}$, for some normal
form~$\ExtCtx{\EmptyCtx}{\BoxTy{\BoolTy}} \vdashNf m : \BoolTy$.
If $m$ is either $\trueTm$ or $\falseTm$, then $\lamTm{x.\, m}$ must
be a constant function and we are done.
Otherwise, it must be some normal
form~$\ExtCtx{\EmptyCtx}{\BoxTy{\BoolTy}} \vdashNf \ifteTm{n,m_1,m_2}
: \BoolTy$ with a neutral~$n : \BoolTy$ either in context~$\EmptyCtx$
or in context~$\ExtCtx{\EmptyCtx}{\BoxTy{\BoolTy}}$.
Such a neutral could either be of shape~$\unboxTm{n'}$
or~$\appTm{n'',m'}$ for some neutrals~$n'$ and~$n''$.
However, this is impossible, since the
context of the neutral~$\unboxTm{n'}$ must contain a lock, and neither the
context~$\EmptyCtx$ nor the context~$\ExtCtx{\EmptyCtx}{\Box{\BoolTy}}$ do.
The existence of $n''$ can also be similarly dismissed by appealing to
the definition of neutrals.

\paragraph{Discussion} Observe that not all Fitch\hyp{}style calculi
are well\hyp{}suited for interpreting the type~$\BoxTy{A}$ as a
secret, because noninterference might not hold.
In \ISFourC, the term~$\lamTm{x.\, \unboxTm{x}} : \BoxTy{A} \FunTy A$
(axiom~\TA) is well\hyp{}typed but leaks the secret~$x$, thus
breaking noninterference.
The validity of the interpretation of $\BoxTy{A}$ as a secret depends
on the calculus under consideration and the axioms it exhibits.

\subsection{Partial Evaluation}\label{sec:staged}

\Textcite{DaviesP96,DaviesP01} present a modal type system for staged
computation based on \ISFour.
In their system, the type~$\BoxTy{A}$ represents \emph{code} of
type~$A$ that is to be executed at a later stage, and the axioms of
\ISFour correspond to operations that manipulate code.
The axiom~$\KA : \BoxTy{(A \FunTy B)} \FunTy (\BoxTy{A} \FunTy \BoxTy{B})$
corresponds to substituting code in code, $\TA : \BoxTy{A} \FunTy A$
to evaluating code, and $\FourA : \BoxTy{A} \FunTy \BoxTy{\BoxTy{A}}$
to further delaying the execution of code to a subsequent stage.
A desired property of this type system is that code must only
depend on code, and thus the term~$\lamTm{x : A.\,\boxTm{x}}$
must be ill\hyp{}typed.

Although \ISFourC exhibits the desired properties of a type system for
staging, its equational theory in \cref{fig:is4c-conversion} does not
reflect the semantics of staged computation.
For example, the result of normalizing the
term~$\boxTm{(2 * \unboxTm{(\boxTm{3})})}$ in \ISFourC extended with
natural number literals and multiplication is $\boxTm{6}$.
While the result expected from reducing it in accordance with
\citeauthor{DaviesP01}'s operational semantics is $\boxTm{(2 * 3)}$.
The equational theory of Fitch\hyp{}style calculi in general do not
take into account the occurrence of a term (such as the literal~$3$)
under $\boxTm$, while this is crucial for \citeauthor{DaviesP01}'s
semantics.
We return to this discussion at the end of this
\lcnamecref{sec:staged}.

If we restrict our attention to a special case of staged computation
in partial evaluation~\parencite{JonesGS93}, however,
the semantics of Fitch\hyp{}style calculi are better suited.
In the context of partial evaluation, the type~$\BoxTy{A}$ represents
a \emph{dynamic} computation of type~$A$ that must be executed at
runtime, and other types represent \emph{static} computations.
Static and dynamic are also known as binding\hyp{}time annotations,
and they are used by a partial evaluator to evaluate all static
computations.

In the term~$\boxTm{(2 * \unboxTm{(\boxTm{3})})}$, we consider
the literal~$3$ to be annotated as dynamic since it occurs under
$\boxTm$.
The construct~$\unboxTm$ strips this annotation and brings it back to
static.
The multiplication of static subterms~$2$ and~$\unboxTm{(\boxTm{3})}$
is however considered annotated dynamic since it itself occurs under
$\boxTm$.
As a result, a partial evaluator that respects these annotations does
not perform the multiplication and specializes the term to
$\boxTm{(2 * 3)}$\mdash{}which matches the result of evaluating with
\citeauthor{DaviesP01}'s staging semantics.
Observe that the same partial evaluator would specialize the
expression~$2 * \unboxTm{(\boxTm{3})}$ to $6$ since the multiplication
does not occur under $\boxTm$ and is thus considered to be annotated
static.

The goal of a partial evaluator is to optimize runtime execution of a
program by eagerly evaluating as many static computations as possible
and yielding an optimal dynamic program.
The term~$\boxTm{6}$ is more optimized than the term~$\boxTm{(2 * 3)}$
since the evidently static multiplication has also been evaluated.
Normalization in a Fitch\hyp{}style calculus yields the former result,
and the gain in optimality can be seen as a form of
\emph{binding\hyp{}time improvement}~\parencite{JonesGS93} that is
performed automatically during normalization.

In this \lcnamecref{sec:staged}, we extend \IKC with natural number
literals and multiplication (denoted \IKCNat), and extend the \NbE
model of \IKC to \IKCNat.
We use \IKC as the base calculus since the other axioms are not needed
in the context of partial evaluation~\parencite{DaviesP96,DaviesP01}.
The resulting normalization function yields an optimal partial
evaluator for \IKCNat.
In partial evaluation, as with staging in general, we desire that a
term~$\lamTm{x : \NatTy.\,\boxTm{x}}$ be disallowed, since a runtime
execution of a dynamic computation must not have a static dependency.
While this term is already ill\hyp{}typed in \IKCNat, we prove a kind
of binding\hyp{}time correctness property for \IKCNat that implies
that \emph{no} term equivalent to $\lamTm{x : \NatTy.\,\boxTm{x}}$ can
exist.

\paragraph{Extension with Natural Number Literals and Multiplication}
We extend \IKC with a type~$\NatTy$, a construct~$\liftTm$ for
including natural number literals, and an operation~$\mulTm$ for
multiplying terms of type~$\NatTy$\mdash{}as described in
\cref{fig:natc-full}.
\begin{figure}[H]
  \begin{align*}
    \Ty\quad A, B \Coloneqq \ldots\ |\ \NatTy & &
    \Ctx\quad \Gamma \Coloneqq \ldots
  \end{align*}
  \begin{mathpar}
    \inferrule[$\NatTy$\nbhyp{}Lift]{ }[k \in \Nats]{%
      \Gamma \vdash \liftTm{k} : \NatTy
    }\label{rule:Nat-lift/NatC}

    \inferrule[$\NatTy$\nbhyp{}Mul]{%
      \Gamma  \vdash t_1 : \NatTy\\
      \Gamma  \vdash t_2 : \NatTy
    }{%
      \Gamma \vdash t_1 \mulTm t_2 : \NatTy
    }\label{rule:Nat-mul/NatC}%
  \end{mathpar}
  \caption{Types, contexts, intrinsically\hyp{}typed terms of
    \IKCNat (omitting the unchanged rules of \cref{fig:ikc-syn-full})}
  \label{fig:natc-full}
\end{figure}

We extend the equational theory of \IKC with some rules such as
$\liftTm{k_1} \mulTm \liftTm{k_2} \thyeq \liftTm{(k_1 * k_2)}$ (for
natural numbers~$k_1$ and~$k_2$),
$\liftTm{0} \mulTm t \thyeq \liftTm{0}$,
$t \thyeq \liftTm{1} \mulTm t$,
$t \mulTm \liftTm{k} \thyeq \liftTm{k} \mulTm t$,
\todowarn{}
\etc.
The normal forms of \IKCNat include those of \IKC in addition to the
following.
\begin{mathpar}
  \inferrule[Nf/$\NatTy$\textsubscript{1}]{}{%
    \Gamma \vdashNf \liftTm{0} : \NatTy
  }\label{rule:Nat-mulNf/NatC-1}

  \inferrule[Nf/$\NatTy$\textsubscript{2}]{%
    \Gamma  \vdashNe n_1 : \NatTy \\
    \dots \\
    \Gamma  \vdashNe n_j : \NatTy
  }[k \in \Nats \setminus \set{ 0 }]{%
    \Gamma \vdashNf \liftTm{k} \mulTm n_1 \mulTm \dotsb \mulTm n_j : \NatTy
  }\label{rule:Nat-mulNf/NatC-2}%
\end{mathpar}
The normal form~$\liftTm{k} \mulTm n_1 \mulTm \dotsb \mulTm n_j$
denotes a multiplication of a nonzero literal with a sequence of
neutrals of type~$\NatTy$, which can possibly be empty.
The term~$\boxTm{(2 * \unboxTm{(\boxTm{3})})}$ from earlier can be
represented in \IKCNat as
$\boxTm{(\liftTm{2} \mulTm \unboxTm{(\boxTm{(\liftTm{3})})})}$, and its
normal form as $\boxTm{(\liftTm{6})}$.
To extend the \NbE model for \IKC to natural number literals and
multiplication, we use the interpretation presented by
\textcite{ValliappanRL21} for normalizing arithmetic expressions.
Omitting the rule~$\liftTm{0} \mulTm t \thyeq \liftTm{0}$, this
interpretation also resembles the one constructed systematically in
the framework of \textcite{YallopGK18} for
commutative monoids.

\paragraph{Proving Binding\hyp{}Time Correctness} Binding\hyp{}time
correctness for a term~$\EmptyCtx \vdash f : \NatTy \FunTy \BoxTy{\NatTy}$
can be stated similar to noninterference: it must be the case that
$\EmptyCtx \vdash \appTm{f,u_1} \thyeq \appTm{f,u_2} :
\BoxTy{\NatTy}$ for any two
arguments~$\EmptyCtx \vdash u_1, u_2 : \NatTy$.
The satisfaction of this property implies that no well\hyp{}typed term
equivalent to $\lamTm{x : \NatTy.\,\boxTm{x}}$ exists, since
applying it to different arguments would yield different results.
As before with noninterference, we can prove this property by case
analysis on the possible normal forms of $f$.
A normal form of $f$ is either of the form~$\lamTm{x.\, \boxTm{(\liftTm{0})}}$
or~$\lamTm{x.\, \boxTm{(\liftTm{k} \mulTm n_1 \mulTm \dotsb \mulTm n_j)}}$ for some
natural number~$k$ and neutrals~$n_1$,\ldots,~$n_j$ of type~$\NatTy$ in context~$\ExtCtx{\ExtCtx{\EmptyCtx}{\NatTy}}{\lockCtx}$.
In the former case, we are done immediately since
$\lamTm{x.\, \boxTm{(\liftTm{0})}}$ is a constant function that
evidently satisfies the desired criterion.
In the latter case, we observe by induction
that no such neutrals~$n_i$ exist, and hence $f$ must be equivalent to the
function~$\lamTm{x.\, \boxTm{(\liftTm{k})}}$, which is also constant.

As a part of binding\hyp{}time correctness, we may also desire that nonconstant
terms~$\BoxTy{A} \FunTy A$ like $\lamTm{x : \BoxTy{A}.\,\unboxTm{x}}$ be disallowed since a
static computation must not have a dynamic dependency.
This can also be shown by following an argument similar to the proof
of noninterference in \cref{sec:ifc}.

\paragraph{Discussion}
The operational semantics for staged computation is given by
\citeauthor{DaviesP01} via translation to a dual\hyp{}context calculus
for \ISFour, where evaluation under the introduction rule~$\boxTm$ for $\BoxTy$
is disallowed.
While it is possible to implement a normalization function for
\ISFourC that does not normalize under $\boxTm$, this then misses
certain reductions that \emph{are} enabled by the translation.
For instance, the term~$\boxTm{(2 \mulTm \unboxTm{(\boxTm{3})})}$ is already in normal
form if we simply disallow normalization under $\boxTm$, while
the translation ensures the reduction of $\unboxTm{(\boxTm{3})}$ by
reducing the term to $\boxTm{(2 \mulTm 3)}$.
This mismatch, in addition to the lack of a model for their system,
makes the applicability of Fitch\hyp{}style calculi for staged
computation unclear.

\section{Related and Further Work}

\paragraph{Fitch\texorpdfstring{\hyp{}}{-}Style Calculi}
Fitch\hyp{}style modal type systems~\parencite{Borghuis94,MartiniM96}
adapt the proof methods of Fitch\hyp{}style natural deduction
systems for modal logic.
In a Fitch\hyp{}style natural deduction system, to eliminate a
formula~$\BoxTy{A}$, we open a so\hyp{}called strict subordinate proof
and apply an ``import'' rule to produce a formula~$A$.
Fitch\hyp{}style lambda calculi achieve a similar effect, for
example in \IKC, by adding a $\lockCtx$ to the context.
To introduce a formula~$\BoxTy{A}$, on the other hand, we close a
strict subordinate proof, and apply an ``export'' rule to a
formula~$A$\mdash{}which corresponds to removing a $\lockCtx$ from the
context.
In the possible\hyp{}world reading, adding a $\lockCtx$ corresponds to travelling
to a future world, and removing it corresponds to returning to
the original world.

The Fitch\hyp{}style calculus~\IKC was presented for the logic~\IK by
\textcite{Borghuis94} and \textcite{MartiniM96}, and later
investigated further by \textcite{Clouston18}.
\citeauthor{Clouston18} showed that $\lockCtx$ can be interpreted as
the left adjoint of $\BoxTy$, and proves a completeness result for a
term calculus that extends \IKC with a type former~$\lockTy$
that internalizes $\lockCtx$.
The extended term calculus is, however, somewhat unsatisfactory since
the normal forms do not enjoy the subformula property.
Normalization was also considered by \citeauthor{Clouston18}, but only
with \cref{rule:box-beta} and not \cref{rule:box-eta}.
The normalization result presented here considers both rules,
and the corresponding completeness result achieved using the \NbE
model does not require the extension of \IKC with $\lockTy$.
The decidability result that follows for the complete equational
theory of \IKC also appears to have been an open problem
prior to our work.

For the logic~\ISFour, there appear to be several possible
formulations of a Fitch\hyp{}style calculus, where the difference has
to do with the definition of the rule~\labelcref{rule:unboxTm/ISFourC-0}.
One possibility is to define $\unboxTm$ by explicitly recording
the context extension as a part of the term former.
\Textcite{DaviesP96,DaviesP01} present such a system where they annotate the
term former~$\unboxTm$ as $\unboxTm_{n}$ to denote the number of
$\lockCtx$s.
Another possibility is to define $\unboxTm$ without any explicit
annotations, thus leaving it ambiguous and to be inferred from a
specific typing derivation.
Such a system is presented by \textcite{Clouston18},
and also discussed by \citeauthor{DaviesP01}.
In either formulation terms of type~$\BoxTy{A} \FunTy A$ (axiom~$\TA$)
and~$\BoxTy{A} \FunTy \BoxTy{\BoxTy{A}}$ (axiom~$\FourA$) that satisfy
the comonad laws are derivable.
As a result, both formulations exhibit the logical
equivalence~$\BoxTy{\BoxTy{A}} \Leftrightarrow \BoxTy{A}$.
The primary difference lies in whether this logical equivalence can
also be shown to be an isomorphism, \ie{} whether the semantics of the
modality~$\BoxTy$ is a comonad which is also \emph{idempotent}.
In \citeauthor{Clouston18}'s categorical semantics the
modality~$\BoxTy$ is interpreted by an idempotent comonad.
The \ISFourC calculus presented here falls under the former category,
where we record the extension explicitly using a premise instead of
an annotation.
\Textcite{GratzerSB19} present yet another possibility that reformulates
the system for \ISFour in \textcite{Clouston18}.
They further extend it with dependent types, and also prove
a normalization result using \NbE with respect to an equational theory that includes
both \cref{rule:box-beta} and \cref{rule:box-eta}.
Although their approach is semantic in the sense of using \NbE, their semantic
domain has a very syntactic flavour
\parencite[Section~3.2]{GratzerSB19} that obscures the elegant
possible\hyp{}world interpretation.
For example, it is unclear as to how their \NbE algorithm
can be adapted to minor variations in the syntax such as in
\IKC, \IKFourC and \ITC{}\mdash{}a solution to which is at the very core
of our pursuit.
This difference also has to do with the fact that they are
interested in \NbE for type\hyp{}checking (also called ``untyped'' or
``defunctionalized'' \NbE),
while we are interested in \NbE for well\hyp{}typed terms
(and thus ``typed'' \NbE), which is better suited
for studying the underlying models.
Furthermore, we also avoid several complications that arise
in accommodating dependent types in a Fitch\hyp{}style calculus,
which is the main goal of their work.

\citeauthor{DaviesP01} present their calculus for \ISFour using a
stack
of contexts%
, which they call ``Kripke\hyp{}style'', as opposed to the single
Fitch\hyp{}style context with a first\hyp{}class delimiting
operator~$\lockCtx$.
The elimination rule~$\unboxTm_{n}$ for $\BoxTy$ in the
Kripke\hyp{}style calculus for \ISFour is indexed by an arbitrary
natural number~$n$ specifying the number of stack frames the rule adds
to the context stack of its premise.
This index~$n$ corresponds to the modal accessibility premise of the
Fitch\hyp{}style $\unboxTm$ rule presented in \cref{fig:is4-full}.
As in the Fitch\hyp{}style presentation, Kripke\hyp{}style calculi
corresponding to the other logics~\IK,~\IT and~\IKFour can be
recovered by restricting the natural numbers~$n$ for which the
$\unboxTm_{n}$ rule is available.
\Textcite{HuP22} present a normalization by evaluation proof for the
Kripke\hyp{}style calculi for all four logics~\IK,~\IT,~\IKFour,
and~\ISFour.
Their solution has a syntactic flavour similar to
\textcite{GratzerSB19} and also does not leverage the
possible\hyp{}world semantics.
Furthermore, their proof is given for a single parametric system that
encompasses the modal logics of interest, which need not be possible
when we consider further modal axioms such as
$\RA : A \FunTy \BoxTy{A}$.

\paragraph{Possible\texorpdfstring{\hyp{}}{-}World Semantics for Fitch\texorpdfstring{\hyp{}}{-}Style Calculi}
Given that Fitch\hyp{}style natural deduction for modal
logic has itself been motivated by possible\hyp{}world
semantics, it is only natural that Fitch\hyp{}style calculi
can also be given possible\hyp{}world semantics.
It appears to be roughly understood that the $\lockCtx$ operator
models some notion of a past world, but this has not been\mdash{}to
the best of our knowledge\mdash{}made precise with a concrete
definition that is supported by a soundness and completeness result.
As noted earlier, this requires a minor refinement of the
frame conditions that define possible\hyp{}world models for
intuitionistic modal logic given by \textcite{BozicD84}.

\paragraph{Dual\texorpdfstring{\hyp{}}{-}Context Calculi}
Dual\hyp{}context calculi~\parencite{PfenningD01,DaviesP96,DaviesP01,Kavvos20} provide
an alternative approach to programming with the necessity modality using
judgements of the form~$\Delta;\Gamma \vdash A$ where $\Delta$ is
thought of as the modal context and $\Gamma$ as the usual (or ``local'') one.
As opposed to a ``direct'' eliminator as in Fitch\hyp{}style calculi,
dual\hyp{}context calculi feature a pattern\hyp{}matching eliminator
formulated as a let\hyp{}construct.
The let\hyp{}construct allows a type $\BoxTy{A}$ to be eliminated into
an arbitrary type~$C$, which induces an array of
commuting conversions in the equational theory to attain normal
forms that obey the subformula property.
Furthermore, the inclusion of an $\eta$\nbhyp{}law for the
$\BoxTy$ type former complicates the ability to produce a unique normal form.
Normalization (and, more specifically, \NbE) for a pattern\hyp{}matching
eliminator\mdash{}while certainly achievable\mdash{}is a much
more tedious endeavour, as evident from the work on
normalizing sum types~\parencite{AltenkirchDHS01,Lindley07,AbelS19},
which suffer from a similar problem.
Presumably for this reason, none of the existing normalization results
for dual\hyp{}context calculi consider the $\eta$\nbhyp{}law.
The possible\hyp{}world semantics of dual\hyp{}context calculi
is also less apparent, and it is unclear how \NbE models
can be constructed as instances of that semantics.

\paragraph{Multimodal Type Theory (\texorpdfstring{\MTT}{MTT})}

\textcite{GratzerKNB20} present a multimodal dependent type theory
that for every choice of mode theory yields a dependent type theory
with multiple interacting modalities.
In contrast to Fitch\hyp{}style calculi, their system features a
variable rule that controls the use of variables of modal type in
context.
Further, the elimination rule for modal types is formulated in the
style of the let\hyp{}construct for dual\hyp{}context calculi.
In a recent result, \Textcite{Gratzer21} proves normalization
for multimodal type theory.
In spite of the generality of multimodal type theory, it is worth noting
that the normalization problem for Fitch\hyp{}style calculi,
when considering the full equational theory, is not a special
case of normalization for multimodal type theory.

\paragraph{Further Modal Axioms}
The possible\hyp{}world semantics and \NbE models presented
here only consider the logics~\IK,~\IT,~\IKFour and~\ISFour.
We wonder if it would be possible to extend the ideas presented
here to further modal axioms such as $\RA : A \FunTy \BoxTy{A}$
and $\GLA : \BoxTy(\BoxTy{A} \FunTy A) \FunTy \BoxTy{A}$,
especially considering that the calculi may differ
in more than just the elimination
rule for the $\BoxTy$ type.

\begin{acks}
  We would like to thank Andreas Abel, Thierry Coquand, and
  Graham Leigh for their feedback on earlier versions of this work.
  We would also like to thank the anonymous referees of both
  the paper and the artifact for their valuable comments and helpful
  suggestions.

  This work is supported by the \grantsponsor{SSF}{Swedish Foundation
  for Strategic Research~(SSF)}{https://strategiska.se} under the
  projects Octopi~(Ref.~\grantnum{SSF}{RIT17-0023R}) and
  WebSec~(Ref.~\grantnum{SSF}{RIT17-0011}).
\end{acks}

\printbibliography

\listoftodos
\end{document}